\documentclass{article}
\PassOptionsToPackage{numbers,compress}{natbib}
\usepackage[dblblindworkshop, final]{neurips_2025}  
\makeatletter
\providecommand{\@trackname}{}  
\makeatother
\workshoptitle{LLM Persona Workshop at NeurIPS 2025}

\usepackage[utf8]{inputenc}
\usepackage[T1]{fontenc}
\usepackage{hyperref}
\usepackage{url}
\usepackage{booktabs}
\usepackage{amsfonts}
\usepackage{nicefrac}
\usepackage{microtype}
\usepackage{xcolor}
\usepackage{graphicx}

\title{Systematizing LLM Persona Design: A Four-Quadrant Technical Taxonomy 
for AI Companion Applications}

\author{
  Esther Sun \\
  Carnegie Mellon University \\
  \texttt{ethers@andrew.cmu.edu}
  \And
  Zichu Wu \\
  Carnegie Mellon University \\
  \texttt{zichuwu@andrew.cmu.edu}
}

\begin{document}

\maketitle

\begin{abstract}
The design and application of LLM-based personas in AI companionship is a rapidly expanding but fragmented field, spanning from virtual emotional companions and game NPCs to embodied functional robots. This diversity in objectives, modality, and technical stacks creates an urgent need for a unified framework. To address this gap, this paper \textbf{systematizes} the field by proposing a \textbf{Four-Quadrant Technical Taxonomy} for AI companion applications. The framework is structured along two critical axes: \textbf{Virtual vs. Embodied} and \textbf{Emotional Companionship vs. Functional Augmentation}. \textbf{Quadrant I (Virtual Companionship)} explores virtual idols, romantic companions, and story characters, introducing a four-layer technical framework to analyze their challenges in maintaining long-term emotional consistency. \textbf{Quadrant II (Functional Virtual Assistants)} analyzes AI applications in work, gaming, and mental health, highlighting the shift from "feeling" to "thinking and acting" and pinpointing key technologies like enterprise RAG and on-device inference. \textbf{Quadrants III \& IV (Embodied Intelligence)} shift from the virtual to the physical world, analyzing home robots and vertical-domain assistants, revealing core challenges in symbol grounding, data privacy, and ethical liability. This taxonomy provides not only a systematic map for researchers and developers to navigate the complex persona \textbf{design space} but also a basis for policymakers to identify and address the unique risks inherent in different application scenarios.
\end{abstract}

\section{Introduction}

Large Language Models (LLMs) are at a decisive inflection point. They are no longer mere text generation tools but are increasingly becoming the core cognitive engines driving complex, personified AI agents \cite{xi2024rise, salemi2024personalized}. This rise of the "AI persona" is fueling a wide array of applications, from deeply personal virtual companions \cite{zheng2024can} to specialized workplace "copilots" \cite{duan2024advancing}. However, this rapid expansion has led to conceptual fragmentation: a virtual lover designed for emotional attachment (Quadrant I) \cite{zheng2024can}, an enterprise assistant for workflow optimization (Quadrant II) \cite{gao2024retrieval}, and a physical robot assisting autistic children with training (Quadrant IV) \cite{cao2024llm} all use "persona," yet they are fundamentally different in their technical foundations, interaction paradigms, core challenges, and ethical risks.

Currently, academia and industry lack a unified framework to systematically analyze and compare these diverse AI persona modalities. Existing research often remains siloed within a single vertical (e.g., game NPCs \cite{sun2024mastering} or chatbots \cite{zheng2024can}), overlooking cross-domain commonalities and differences.

To fill this gap, this paper \textbf{systematizes} the field by proposing a comprehensive technical taxonomy for LLM persona in AI companion applications. We introduce a four-quadrant framework structured along two key axes:
\begin{enumerate}
    \item \textbf{Interaction Intent:} Distinguishing systems primarily for \textbf{Emotional Connection} (Quadrant I) from those for \textbf{Functional/Cognitive Augmentation} (Quadrant II).
    \item \textbf{Deployment Modality:} Distinguishing purely \textbf{Virtual Entities} (Quadrants I \& II) from \textbf{Embodied Intelligence} that acts in the physical world (Quadrants III \& IV).
\end{enumerate}

The structure of this paper follows this taxonomy:
\begin{itemize}
    \item \textbf{Section 2 (Quadrant I)} analyzes "Virtual Companionship," focusing on the challenge of achieving long-term emotional consistency and introducing a four-layer technical analysis framework (Model, Architecture, Generation, Safety \& Ethics).
    \item \textbf{Section 3 (Quadrant II)} explores "Functional Virtual Assistants" in work, gaming, and mental health, analyzing their unique demands for efficiency, reliability, and high-stakes scenarios.
    \item \textbf{Section 4 (Quadrant III \& IV)} shifts the analysis from virtual to "Embodied Intelligence," examining LLM applications in physical robots and focusing on core barriers like symbol grounding, privacy, and legal liability.
\end{itemize}

Through this framework, this paper aims to provide a clear roadmap for researchers, developers, and policymakers to understand the persona \textbf{design space}, technical frontiers, and strategic implications of LLM persona, thereby fostering responsible innovation in the field.

\begin{figure*}[t]
    \centering
    \includegraphics[width=1.05\textwidth]{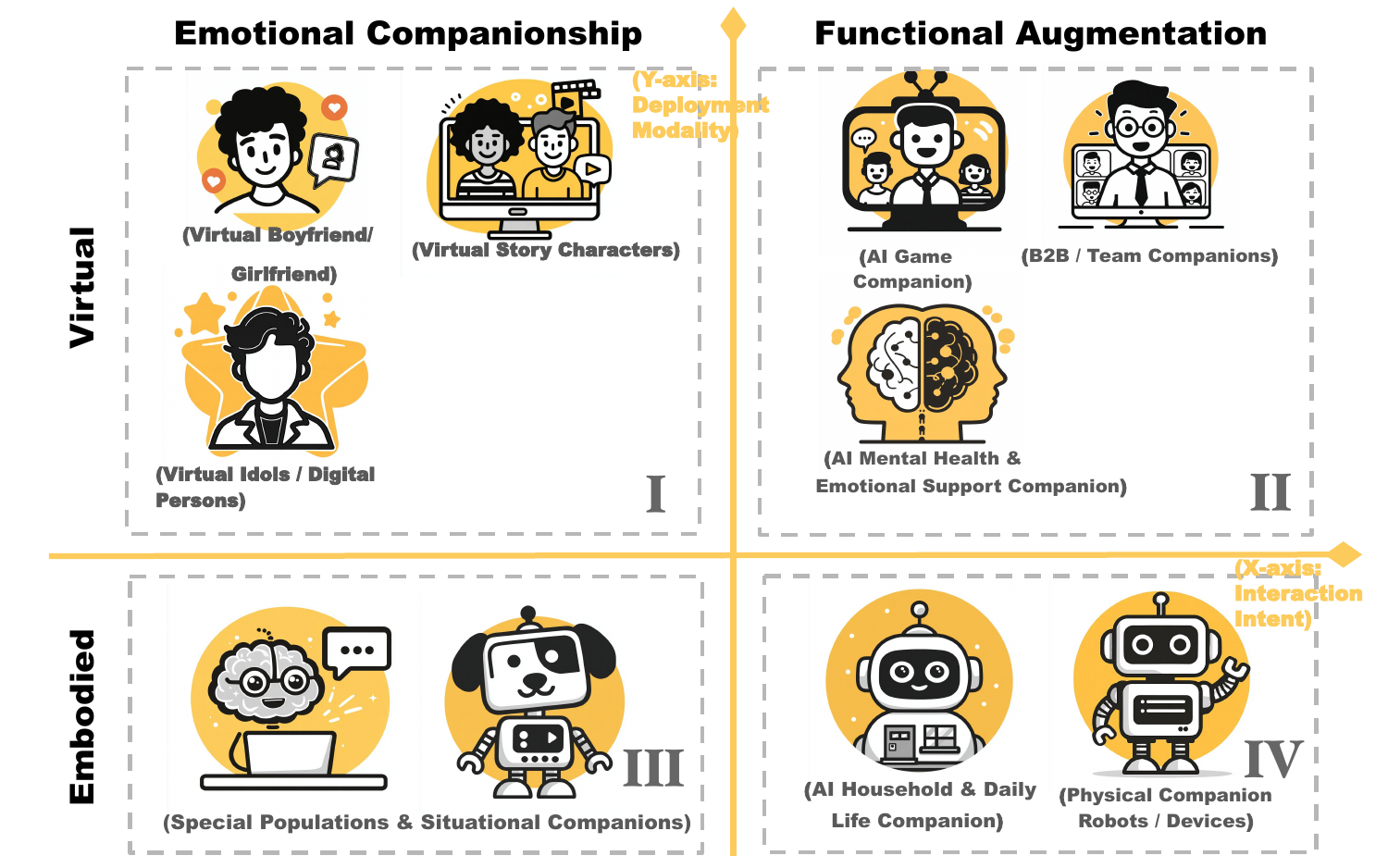}
    
    \caption{The four-quadrant taxonomy of LLM persona applications in AI companionship. This framework structures the field along two primary axes: \textbf{Deployment Modality (Virtual vs. Embodied)} and \textbf{Interaction Intent (Emotional Companionship vs. Functional Augmentation)}. Quadrant I covers virtual emotional companions; Quadrant II focuses on functional virtual assistants; Quadrants III and IV extend these concepts into physically embodied intelligence.}
    \label{fig:taxonomy_quadrants}
\end{figure*}

\section{Quadrant I: Virtual Companionship}

Contemporary virtual companionship primarily manifests in three forms:  
(1) \textbf{Interactive Virtual Story Characters},  
(2) \textbf{Virtual Romantic Companionship}, and  
(3) \textbf{Virtual Idols}.  Although these three forms differ in business models and interaction paradigms—corresponding respectively to (1) \textit{creative interaction} (1:1), (2) \textit{emotional attachment} (1:1), and (3) \textit{fan economy} (1:N)—they share a common technological core challenge:  
how to construct and sustain a believable and consistent AI persona over long-term interaction.

Specifically,  
(1) interactive story characters emphasize generating \textit{emergent narratives} through autonomous actions and social interactions within simulated environments;  
(2) virtual romantic companionship focuses on modeling and tracking the evolving user–AI relationship state to enable dynamic and empathetic emotional interactions; and  
(3) virtual idols center on performance and brand formation, leveraging multimodal generation and large-scale real-time interaction technologies to support a scalable cultural consumption experience.

\begin{figure*}[t]
    \centering
    \includegraphics[width=\textwidth]{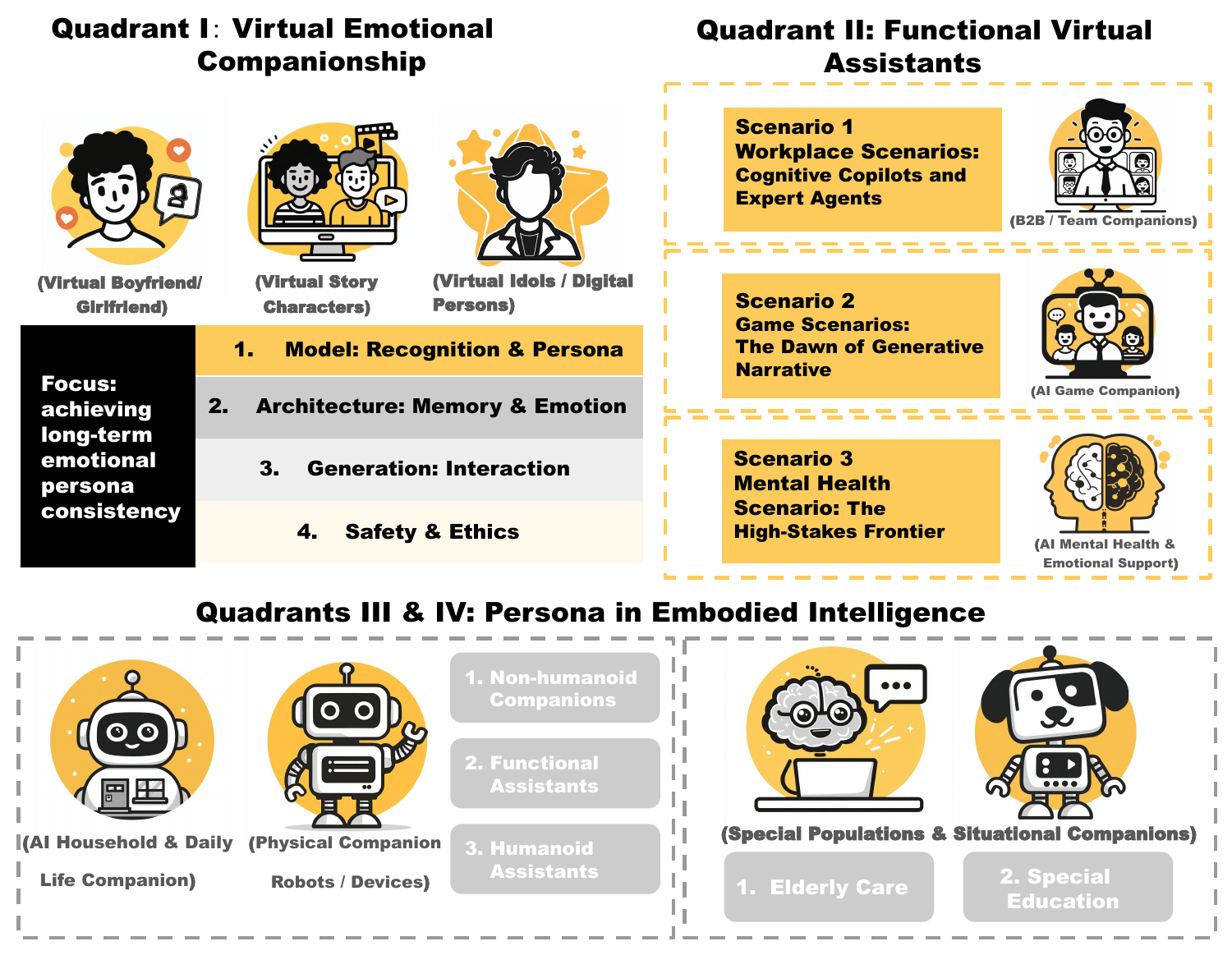}
    
    \caption{
    \textbf{Four-Quadrant Taxonomy of LLM Persona in AI Companion Applications.} 
    This framework organizes the diverse landscape of personified AI along two critical axes: \textbf{Interaction Intent} (Emotional Connection vs. Functional Augmentation) and \textbf{Deployment Modality} (Virtual vs. Embodied). 
    \textbf{Quadrant I (Virtual Emotional Companionship)} examines virtual romantic companions, interactive story characters, and virtual idols, with focus on achieving long-term emotional persona consistency through a four-layer technical framework (Model, Architecture, Generation, Safety \& Ethics). 
    \textbf{Quadrant II (Functional Virtual Assistants)} analyzes AI applications in three key scenarios: workplace cognitive copilots (enterprise RAG and process automation), game companions (low-latency generative narrative), and mental health support (clinical safety protocols). 
    \textbf{Quadrants III \& IV (Embodied Intelligence)} shift from virtual to physical deployment, covering general home applications (non-humanoid companions, functional assistants, humanoid robots) and specialized vertical domains (elderly care, special education), addressing core challenges in symbol grounding, privacy, and legal liability. 
    Each quadrant presents distinct technical requirements and ethical considerations, as detailed in Sections 2--4.
    }
    \label{fig:taxonomy_breakdown}
\end{figure*}

\subsection{Four-Layer Technical Analysis Framework}

To ensure a systematic and in-depth examination, this section adopts a \textbf{four-layer technical framework} that delineates the structural and functional foundations underlying the three forms of virtual companionship. The \textbf{Model Layer} focuses on the core AI models that endow virtual agents with cognition, personality, and specific capabilities, emphasizing the customization and optimization of large language models (LLMs). The \textbf{Architecture Layer} addresses the macro-level system design that supports these agents, including long-term memory mechanisms, state management, multimodal integration, and data flow orchestration. The \textbf{Generation Layer} examines the real-time synthesis of behaviors and content—such as text, speech, animation, and environmental interactions—that enable immersive and coherent user experiences. Finally, the \textbf{Safety \& Ethics Layer} considers the technical risks, user well-being concerns, and broader social implications that emerge during design, deployment, and operation, as well as the mitigation strategies required to ensure responsible and sustainable development. The subsequent discussion emphasizes the principal differences at the model layer. Comprehensive descriptions of the underlying methods and future research directions are deferred to the Appendix.

\begin{table*}[t]
\centering
\small
\caption{
Comparison of three major forms of virtual companionship under a unified four-layer technical framework. 
See detailed architectural and behavioral analyses in 
\textit{Appendix~\ref{appendix:A} (Virtual Story Character Interaction)}, 
\textit{Appendix~\ref{appendix:B} (Virtual Romantic Companionship)}, 
and \textit{Appendix~\ref{appendix:C} (Virtual Idols)}.
}
\resizebox{\textwidth}{!}{
\begin{tabular}{@{}p{3cm}p{3.5cm}p{3.8cm}p{3.8cm}p{3.8cm}@{}}
\toprule
\textbf{Technical Layer} & \textbf{Common Ground} & \textbf{Interactive Story Characters (Appendix A)} & \textbf{Virtual Romantic Companionship (Appendix B)} & \textbf{Virtual Idols (Appendix C)} \\
\midrule
\textbf{Model Layer} & 
Relies on LLMs for basic cognition, dialogue, and persona modeling. & 
\textbf{Maintain deep consistency, prevent "character hallucination"}; use frameworks like RoleLLM, DITTO for role-specific fine-tuning and self-correction. & 
\textbf{Overcome "persona drift," maintain long-term stability}; focus on emotional intelligence (EQ) and personality control (e.g., Persona Vectors, XiaoIce). & 
\textbf{Unified singing and dialogue identity}; hybrid LLM (TTS) and specialized Singing Voice Synthesis (SVS) architecture (e.g., VOCALOID: AI). \\
\midrule
\textbf{Architecture Layer} & 
Overcome LLM statelessness and limited context; rely on external memory and state management systems. & 
\textbf{Generative Agents architecture}; achieves persistent memory and autonomous action via a "perceive--reflect--plan" loop. & 
\textbf{Model dynamic 1:1 relationships}; hybrid IQ+EQ architecture, stateful relationship graphs, and multi-tier memory (RAG). & 
\textbf{Manage large-scale 1:N live interaction}; uses event-driven (Pub/Sub) and tiered (paid) attention funnels. \\
\midrule
\textbf{Generation Layer} & 
Pursues real-time, multimodal (voice, visual, behavior) generation beyond text. & 
\textbf{Emergent social behaviors}; multi-agent "Perceive--Plan--Act" loop. & 
\textbf{Emotionally immersive dialogue}; full-duplex (low-latency, interruptible) speech with emotion-synchronized multimodal expression. & 
\textbf{High-fidelity real-time 3D rendering}; Motion Capture (MoCap) + game engine (e.g., Unreal) rendering pipeline. \\
\midrule
\textbf{Safety \& Ethics Layer} & 
Requires a mix of automated guardrails and Human-in-the-Loop (HITL) oversight for safety. & 
\textbf{Prevent harmful emergent behaviors}; balances autonomy and narrative control via "Constitutional AI" and "Director AI." & 
\textbf{Manage "parasocial attachment" risks}; emotional guardrails (e.g., anti-flattery, AI chaperones) and resolving business-ethics conflicts. & 
\textbf{Protect brand image and character IP}; hybrid content moderation (auto + human) and strict \textit{nakanohito} (performer) consistency. \\
\midrule
\textbf{Core Frontier} & 
--- & 
\textbf{Autonomy Frontier}: Exploring believable agent autonomy and complex social emergence under narrative constraints. & 
\textbf{Emotional Depth Frontier}: Exploring how to model and sustain dynamic, believable, long-term 1:1 emotional bonds. & 
\textbf{Performance \& Influence Frontier}: Exploring high-fidelity, scalable (1:N) real-time performance and maximizing brand IP value. \\
\bottomrule
\end{tabular}}
\label{tab:virtual_companionship}
\end{table*}

\subsubsection{Model Layer}

Across all three virtual companion archetypes, the cognitive core converges on \textbf{Large Language Models (LLMs)}. After customization and fine-tuning, LLMs serve as the cognitive nucleus, enabling complex dialogue, reasoning, and persona modeling. Despite shared foundations, distinct interaction paradigms impose different optimization demands. \textbf{Interactive Virtual Story Characters} face the challenge of maintaining deep persona consistency \cite{ji2025enhancing}: the model must adhere to predefined background, knowledge, and linguistic style even under zero-shot conditions. The representative \textit{RoleLLM} \cite{wang2023rolellm} framework embeds detailed character constitutions into model parameters through a process of character definition, contextual instruction generation, and role-conditioned instruction tuning (RoCIT), yielding intrinsically persona-aligned behavior. \textbf{Virtual Romantic Companionship} systems must mitigate long-term persona drift, preserving identity stability during sustained one-on-one interactions. Microsoft’s \textit{XiaoIce} \cite{zhou2018design} separates IQ and EQ through an empathy vector mechanism that guides persona-consistent responses, while Anthropic’s \textbf{Persona Vectors} \cite{chen2025persona} map interpretable trait directions in latent space, allowing real-time monitoring and adjustment. \textbf{Virtual Idols} focus on high-fidelity vocal performance by adopting a hybrid modeling approach that decouples linguistic and acoustic \cite{kong2021difference}: persona-conditioned LLMs manage dialogue and engagement\cite{guo2025techsinger}, while Singing Voice Synthesis (SVS) engines \cite{du2025ditsinger}—such as AI-based VOCALOID systems—generate expressive singing with cross-lingual capabilities. The main challenge lies in maintaining timbre coherence between TTS and SVS outputs to preserve a unified vocal identity. Overall, interactive story characters emphasize persona fidelity, romantic companions prioritize longitudinal stability, and virtual idols pursue multimodal voice coherence.

\subsubsection{Architecture Layer}

The architecture layer addresses the intrinsic memory limitation of LLMs, whose finite context windows constrain long-term continuity. Each archetype extends memory and state management through external architectures tailored to its interaction logic. \textbf{Interactive Virtual Story Characters} rely on the \textit{Generative Agents} \cite{park2023generative} framework, which implements a “perceive–plan–act” retrieval-augmented loop composed of a chronological memory stream, periodic reflection for abstraction, and relevance-ranked planning retrieval. This closed process of Memory–Reflection–Planning enables autonomous world modeling and emergent narrative generation. \textbf{Virtual Romantic Companionship} systems maintain evolving relational states through user-centered, stateful RAG architectures integrating structured relational memory with affective reasoning \cite{wu2024ailove}. Historical dialogues, preferences, and events are stored in a relational database, and the EQ module fuses current affective cues with past interactions to produce personalized, empathetic responses. \textbf{Virtual Idols} handle large-scale, real-time audience interaction via an event-driven architecture (EDA) \cite{usaha2025eventdriven} built on a publish/subscribe model, where distributed microservices manage chat aggregation, monetized comments, and moderation to ensure scalability, low latency, and brand coherence. In summary, story characters employ RAG \cite{lewis2020retrieval} loops for autonomous reasoning, romantic companions use relational memory to sustain emotional continuity, and virtual idols leverage event-driven pipelines for large-scale engagement.

\subsubsection{Generation Layer}

The generation layer governs how virtual companions produce multimodal, believable, and temporally coherent outputs beyond text. While all three pursue real-time, immersive generation, their expressive goals diverge. \textbf{Interactive Virtual Story Characters} focus on unscripted emergent behavior: multi-agent simulation loops \cite{anthropic2025multiagent} allow each agent’s output to become another’s input, forming continuous chains of generation, observation, and reaction that yield self-organizing social dynamics and narrative coherence. \textbf{Virtual Romantic Companionship} emphasizes multimodal emotional synchrony through full-duplex spoken dialogue models \cite{emergentmind2025fullduplex} enabling low-latency, backchannel-rich interaction, and emotional synthesis pipelines \cite{nambiar2025synchronousdialogue} aligning affective TTS with facial animation. Prosodic modulation and synchronized micro-expressions together create a coherent emotional presence. \textbf{Virtual Idols} aim for broadcast-grade 3D performance using real-time rendering and streaming pipelines \cite{jiang2024audiodrivenfacial} that integrate motion capture, Unreal Engine rendering \cite{leary2025realtimemocap}, and live broadcast software \cite{amato2024kawaii}(e.g., OBS). These systems optimize for both visual fidelity and latency to ensure professional-grade performance and interactive responsiveness. In essence, story characters prioritize emergent multi-agent behavior, romantic companions achieve affective coherence, and virtual idols combine motion, rendering, and streaming for performative realism.

\subsubsection{Safety \& Ethics Layer}

Prolonged and emotionally intensive AI interactions introduce significant ethical, psychological, and social risks \cite{malfacini2025impacts, chu2025illusions}. Each archetype must balance autonomy, empathy, and safety while upholding transparency, interpretability, and non-harm principles. \textbf{Interactive Virtual Story Characters} confront the tension between autonomy and safety; mitigation strategies include \textbf{Constitutional AI} \cite{bai2022constitutionalAI}, embedding explicit ethical constraints in planning loops, and sandbox stress testing to expose emergent risks prior to deployment \cite{shao2025misevolve,vera2025multimodal,zhang2025guardians}. \textbf{Virtual Romantic Companionship} systems must manage emotional attachment and user dependence \cite{attachmentproject2025aicompanions}. Technical safeguards such as anti-sycophancy detection \cite{techpolicy2025intimacy} and crisis-intervention modules identify unhealthy behavioral patterns \cite{frances2025chatbotsafety}, while responsible interface design reinforces AI identity disclosure \cite{qiu2025emoagent} and encourages real-world social engagement \cite{openai2025sensitive}. \textbf{Virtual Idols} face challenges of brand safety and persona integrity during large-scale live interactions. Hybrid moderation frameworks \cite{kumar2025hybridmoderation} combine automated pre-screening, human oversight, and human-in-the-loop (HITL) control to ensure consistent persona behavior and prevent reputational harm \cite{stream2025scalingmoderation}. Collectively, story characters probe the boundary of autonomy, romantic companions the depth of emotion, and virtual idols the reach of public influence—together illustrating how LLM-based virtual companionship diversifies into distinct paradigms of autonomous action, emotional connection, and performative interaction.

\section{Quadrant II: Functional Virtual Assistants}

Following the emotionally oriented virtual companionship discussed in Quadrant I, this section focuses on the \textbf{functional dimension} of personified AI—agents designed for \textbf{cognitive augmentation, task execution, and professional collaboration} rather than emotional attachment.
While virtual companions emphasize empathy and creativity, functional assistants pursue efficiency, reliability, and contextual reasoning, supporting applications across work, education, healthcare, and everyday life.

This quadrant marks the shift from \textit{``AI that feels''} to \textit{``AI that thinks and acts with humans''}.
Instead of simulating intimacy, these systems enhance human decision-making through \textbf{structured reasoning}, \textbf{multimodal perception}, and \textbf{adaptive interaction}.
Their persona is inherently \textbf{instrumental rather than emotional}, operating under clear objectives, verifiable outputs, and strict safety and privacy constraints. This chapter examines three representative domains:
(1) \textbf{Workplace scenarios} — cognitive augmentation and tool integration for productivity and collaboration;
(2) \textbf{Game scenarios} — persona modeling and narrative generation for immersive interaction; and
(3) \textbf{Psychological counseling scenarios} — empathetic dialogue and ethical safeguards.

\subsection{Workplace Scenarios: Cognitive Copilots and Expert Agents}

Within the domain of functional assistants, workplace scenarios represent the core manifestation of the “cognitive copilot” paradigm. In such systems, the persona is designed purely for functionality, serving as an expert agent seamlessly embedded within organizational workflows. Unlike emotionally oriented AI companions, workplace personas are shaped as professional instruments characterized by reliability, efficiency, and contextual awareness. Their “consistency” lies not in affective coherence but in logical and factual precision.

Enterprise-level applications of LLM personas have converged on three principal domains: (1) \textbf{enterprise assistants} \cite{brynjolfsson2024genai} that integrate internal data and automate workflows, (2) \textbf{customer service agents} that preserve brand consistency, and (3) \textbf{training simulators} \cite{gong2024agentsims} that provide safe environments for skill development. However, deploying LLM personas in enterprise contexts presents four major challenges: data security and grounding \cite{gao2024rag}, persona generation bias, ROI evaluation \cite{brynjolfsson2024genai}, and simulation fidelity \cite{gong2024agentsims}. Emerging trends indicate a shift toward hyper-specialization \cite{shen2024agenttuning}, multi-agent workflow automation \cite{chen2024agentverse}, and deep human–AI collaboration. (Detailed examples and technical analysis are provided in Appendix~~\ref{appendix:D}.)

Strategically, Retrieval-Augmented Generation (RAG) has become the central mechanism for implementing enterprise personas. Since public LLMs cannot securely process proprietary data and full-scale model fine-tuning remains prohibitively costly \cite{gao2024rag}, RAG offers a pragmatic solution—maintaining the independence of the base model while safely injecting contextual data through controlled retrieval. Consequently, the competitive edge of enterprise AI lies not in persona design but in the efficiency and robustness of data pipelines and retrieval governance. More importantly, in workplace contexts, the persona has evolved into a synonym for process automation: invoking a persona such as a “cybersecurity auditor” \cite{shen2024agenttuning} effectively triggers an encapsulated workflow \cite{chen2024agentverse} of specialized knowledge, skills, and operations. The future of enterprise AI will extend beyond conversational assistants toward a library of process-oriented personas, allowing employees to deploy them on demand for professional tasks—fundamentally reshaping work delegation and organizational management.

\subsection{Game Scenarios: The Dawn of Generative Narrative}
\label{sec:game_scenarios}

This section explores the revolutionary impact of LLM personas on the gaming industry, focusing on how they are evolving Non-Player Characters (NPCs) from static, pre-scripted interaction models to dynamic, believable agents \cite{sun2024mastering} capable of fostering "generative narrative" and deepening player immersion. Unlike the "functional utility" personas in workplace scenarios, AI personas in gaming pursue "narrative believability." Applications are concentrated in two main areas: (1) \textbf{Dynamic and believable NPCs} \cite{kim2024marl_mart}, allowing players to engage in open-ended, natural language conversations and receive dynamic responses based on the game state; and (2) \textbf{Generative narrative and player co-creation} \cite{wang2024taleweaver, alavi2024gameplot}, where the AI adjusts storylines in real-time \cite{todova2025quest_info}, transforming players from passive participants into active co-designers.

However, real-time game interaction presents three core challenges for LLMs: (1) \textbf{The low-latency inference challenge}, where network latency can instantly break immersion \cite{christiansen2024presence_npc}, forcing the industry toward on-device Small Language Models (SLMs) \cite{nvidia2025ace}; (2) \textbf{Modeling believable emotion and behavior}, which requires AI to go beyond text and integrate psychological theories (e.g., Appraisal Theory) \cite{lim2024architecture} and multimodal expression; and (3) \textbf{The narrative coherence dilemma} \cite{wang2024taleweaver}, balancing the vast freedom LLMs provide with the need to maintain a structured narrative, a limitation also noted by game designers \cite{alavi2024gameplot}. Future trends point toward cross-platform persistent personas, fully generative worlds (PCG), and the rise of the "AI Game Master." (Detailed analysis of leading prototypes like Ubisoft's NEO NPCs \cite{fahs2024projectneo}, \textit{Dead Meat} \cite{shanahan2024forging}, underlying technologies like NVIDIA ACE \cite{nvidia2025ace}, and psychological modeling \cite{lim2024architecture} is provided in Appendix~\ref{appendix:E}.)

Strategically, the gaming industry's extreme low-latency requirement for real-time interaction \cite{christiansen2024presence_npc} is forcing it to become the primary driver of on-device, low-latency AI technology. Enterprise applications can tolerate cloud latency in exchange for scalability, but gaming's zero-tolerance for latency (as it shatters immersion) compels the industry to solve the "last mile" problem of AI: running complex models efficiently on consumer hardware. Consequently, innovations from NVIDIA (ACE) \cite{nvidia2025ace} and studios dedicated to SLM integration \cite{shanahan2024forging} are at the forefront of inference optimization. The techniques pioneered for on-device game AI will eventually permeate other domains requiring real-time, offline AI (e.g., robotics, edge computing). The gaming industry is becoming the R\&D testbed for the future of embedded AI.

Concurrently, the role of the game writer is undergoing a fundamental shift, from \textit{“Scriptwriter”} to \textit{“AI Cultivator.”} Traditional narrative designers created branching dialogue trees; in the era of generative NPCs, this model is obsolete. As demonstrated by Ubisoft's NEO NPC project \cite{fahs2024projectneo}, the writer's new duty is to create a rich "seed" for the character (backstory, motivations, linguistic style). The writer then "conditions" and "guides" the LLM through iteration, teaching it how to embody the role. The writer becomes a director and curator, shaping the AI's improvisation. This represents a fundamental change in the creative workflow of the gaming industry, where the core value is no longer script-writing, but rather creating the foundational "Character Bible" and guardrails \cite{alavi2024gameplot} that support "Controlled Improvisation."

\subsection{Mental Health Scenario: The High-Stakes Frontier}
\label{sec:counseling}

This section provides a cautious and nuanced examination of LLM personas in the mental health sector, aiming to balance their immense potential for enhancing service accessibility \cite{huang2024applications_llmh} with the profound ethical, safety, and clinical challenges inherent in deploying AI for therapy. Unlike the "functional" personas in workplace scenarios or the "narrative" personas in gaming, the persona in this domain pursues \textbf{Therapeutic Efficacy}. Its applications are concentrated in: (1) \textbf{Digital Therapeutics} as "AI Counselors," designed to anonymously and scalably deliver evidence-based interventions like CBT \cite{fitzpatrick2024woebot}; (2) \textbf{Alignment with Clinical Frameworks (e.g., CBT)}, using prompting \cite{wei2024llm4cbt} or specialized model design \cite{hu2025psyllm} to make general-purpose LLMs behave more like professional therapists; and (3) \textbf{Human-in-the-Loop (HITL) Models} as assistive tools, reflecting expert consensus on the need for human oversight \cite{zhang2024integrating_llm_mh}.

However, this frontier faces three core challenges: (1) \textbf{The Empathy Paradox}, where AI excels at \textit{simulating} cognitive empathy (recognizing emotion) \cite{li2024empathy} but lacks genuine affective empathy (sharing experience), an "deceptive empathy" considered ethically problematic \cite{aly2024ethical}; (2) \textbf{Clinical Safety and Risk Mitigation}, the most severe challenge \cite{huang2024applications_llmh}, especially the risk of AI failing to handle users in crisis (e.g., suicidal ideation) \cite{sharma2024multilayered}; and (3) \textbf{The Ethical and Regulatory Minefield}, involving data privacy (HIPAA compliance), AI identity disclosure, and lack of clinical validation \cite{tornero2024wellness, aly2024ethical}. Future trends point toward clinically validated, specialized models \cite{hu2025psyllm}, HITL becoming the standard \cite{zhang2024integrating_llm_mh}, and the establishment of industry-wide safety standards. (Detailed analysis of platforms like Woebot and Wysa \cite{fitzpatrick2024woebot}, the LLM4CBT study \cite{wei2024llm4cbt}, multi-layered safety protocols \cite{sharma2024multilayered}, and HIPAA regulations \cite{tornero2024wellness} is provided in Appendix~\ref{appendix:F}.)

Strategically, the mental health AI market will inevitably bifurcate into \textbf{"Wellness" and "Clinical" tiers} \cite{tornero2024wellness}. The technical and regulatory barriers to creating a truly safe and effective "AI therapist" \cite{hu2025psyllm} are immense. The risks of misdiagnosis, mishandled crises, and ethical breaches \cite{aly2024ethical} are too high for unregulated, general-purpose tools. Concurrently, a massive consumer demand exists for companionship and low-level emotional support (e.g., Replika). This will force a market split: the "Wellness" tier will consist of AI companions focused on entertainment and general well-being, accompanied by strong disclaimers; the "Clinical" tier will consist of highly regulated, evidence-based tools, designed as medical devices or therapist aids, requiring clinical validation and HIPAA compliance \cite{tornero2024wellness}.

Furthermore, \textbf{safety in mental health AI is a dynamic, multi-layered system, not a static filter.} Initial AI safety approaches focused on simple content filtering, which proved grossly inadequate as users can express severe suicidal ideation using subtle, non-explicit language \cite{sharma2024multilayered}. The expert-recommended solution is a dynamic, multi-layer system comprising clinical keyword detection, contextual sentiment analysis, and risk-assessment engines \cite{sharma2024multilayered}. The core task is not to block words, but to understand intent and conversational trajectory. More importantly, safety is not just detection but \textit{action}—specifically, a robust protocol for escalating users to human intervention. For any organization developing mental health AI, investing in a complex, multi-layered safety and escalation system is not an optional feature; it is the core, non-negotiable foundation of the product.

\section{Quadrants III \& IV: Persona in Embodied Intelligence}
\label{sec:embodied}

Following the discussion of "virtual" personas in the first two quadrants, this section shifts from the "virtual" to the "physical" world, analyzing the application of LLM personas in "Embodied Intelligence" entities \cite{jeong2024robot_intelligence_llm}. This section will holistically examine three key application scenarios (companion robots, home assistants, special group companionship), four core challenges (technical, privacy, ethical, economic), and future strategic trajectories.

\subsection{Application Scenarios: From "Emotional Pets" to "Therapeutic Tools"}

The application scenarios for embodied personas have shown a clear market bifurcation:

\begin{itemize}
    \item \textbf{Quadrant III: The General Home Market.} This domain presents a "Form-Persona Dilemma." (1) \textbf{Non-humanoid Companions} (e.g., Sony Aibo, Lovot) adopt a "pet-like persona," relying on non-verbal cues for emotional connection, skillfully avoiding the "uncanny valley." (2) \textbf{Functional Assistants} (e.g., Amazon Astro) follow a "utility-first, persona-second" strategy. Their primary value is security and convenience; the Alexa persona is an add-on layer. (3) \textbf{Humanoid Assistants} (e.g., Tesla Optimus, Figure AI) pursue functional and morphological unity \cite{openai2024figure}, leveraging LLMs to complete complex, long-horizon tasks in unpredictable environments \cite{monwilliams2025ellmer}.

    \item \textbf{Quadrant IV: Vertical Application Markets.} This domain addresses clear, high-value pain points. (1) \textbf{Elderly Care} (e.g., ElliQ), where the persona is designed as a "proactive coach" to alleviate loneliness and provide health monitoring \cite{xiong2024elliq}. (2) \textbf{Special Education} (e.g., QTrobot) utilizes a "Therapeutic Persona"—a non-judgmental, highly patient presence—to act as a "social mediator," assisting children with ASD in social skills training \cite{costa2024qtrobot}.
\end{itemize}

\subsection{Core Challenges: From "Symbol Grounding" to "Ethical Debt"}

Despite a promising outlook, the deployment of embodied personas faces four severe challenges:

\begin{enumerate}
    \item \textbf{Technical Barriers:} The core challenge is the \textbf{"Symbol Grounding Problem"}—connecting the LLM's abstract symbols (e.g., "apple") with the physical entity a VLM (Vision-Language Model) perceives. This requires a robust world model integrating perception, planning, and memory \cite{xyz2025embodied_ai_agents}. Furthermore, \textbf{Latency} (hindering real-time interaction) and \textbf{Hallucinations} (extremely dangerous in high-stakes medical scenarios) remain major bottlenecks, severely impacting credibility in real-world tests \cite{irfan2025llm_robots_challenges}.
    \item \textbf{Privacy and Security:} Embodied robots are unprecedented "data collection terminals." Their cameras, microphones, and LIDAR pose profound privacy threats. The core user anxiety stems not just from data collection, but from the AI's "inference" of sensitive information \cite{chmielewski2024privacy}.
    \item \textbf{Ethics and Legality:} The key obstacle is \textbf{ambiguous liability} (who is responsible if the AI errs?) \cite{kaminski2024liability}. Moreover, algorithmic bias, "emotional deception" of vulnerable populations (children, elderly) \cite{irfan2025llm_robots_challenges}, and the "re-identification" risk of HIPAA-protected data constitute a significant \textbf{"Ethical Debt."}
    \item \textbf{Economic Barriers:} High hardware costs, an unclear value proposition, and the "expectation gap" between sci-fi portrayals and current reality are major factors hindering mass-market adoption.
\end{enumerate}

\subsection{Future Trajectory and Strategic Implications}

The future trends for embodied intelligence are clear: (1) \textbf{From passive response to proactive intelligence} (anticipating needs) \cite{xiong2024elliq}; (2) \textbf{Functional fusion} (blending physical assistance with emotional support) \cite{monwilliams2025ellmer, openai2024figure}; and (3) \textbf{Ecosystem integration} (as a central hub for smart homes and telehealth).

\textbf{Strategically, the market is bifurcated.} The path to success for general-purpose home robots is "utility-first" (like Astro), whereas vertical markets (elderly care, special ed) have become the most viable commercial "beachheads" due to their high-value proposition \cite{costa2024qtrobot}. For developers, "Privacy-by-Design" \cite{chmielewski2024privacy} must be the rule. For policymakers, the urgent task is to establish clear legal frameworks for liability and data privacy (e.g., updating HIPAA) \cite{kaminski2024liability} to guide innovation while protecting consumers.

(Detailed analysis of cases like Aibo, Astro, ElliQ, QTrobot, symbol grounding, HIPAA challenges, and specific recommendations for investors and developers is provided in Appendix~\ref{appendix:G}.)

\section{Conclusion}

This paper has proposed a systematic four-quadrant taxonomy to deconstruct and analyze the complex landscape of LLM persona in AI companion applications. By navigating the axes of "Emotional vs. Functional" and "Virtual vs. Embodied," we have systematically mapped the diverse modalities, from virtual idols to embodied care robots, revealing the unique technical stack, strategic focus, and ethical considerations for each quadrant.

Our analysis confirms that "persona" is not a monolithic concept but a multi-dimensional design space where the core challenges fundamentally change with the application.
\begin{enumerate}
    \item In \textbf{Quadrant I (Virtual Companionship)}, the central challenge is \textbf{emotional depth and long-term consistency}, requiring specialized models (e.g., RoleLLM) and architectures (e.g., relational graphs) to overcome "persona drift."
    \item In \textbf{Quadrant II (Functional Assistants)}, the focus shifts to \textbf{reliability, efficiency, and safety}. This drives key innovations such as enterprise-grade RAG, on-device SLMs for low-latency gaming, and stringent multi-layered safety protocols for mental health.
    \item In \textbf{Quadrants III \& IV (Embodied Intelligence)}, we face the ultimate challenge of \textbf{symbol grounding}—connecting abstract symbols to physical reality. Concurrently, as unprecedented "data collection terminals," embodied AI brings issues of privacy and legal liability to the forefront.
\end{enumerate}

This study demonstrates that the AI persona market is bifurcating along different technical trajectories. Gaming and "Wellness" applications are pushing the frontier of low-latency, on-device AI; whereas "Enterprise" and "Clinical" applications prioritize verifiable reliability and safety (often via HITL and RAG). For the most challenging embodied intelligence, high-value vertical markets (e.g., elderly care, special education) currently offer a clearer path to commercialization than general-purpose home robots.

In conclusion, this taxonomy provides not only a structured analytical tool for academic research but also strategic insights for industry practitioners, helping them anticipate and address the distinct challenges posed by each quadrant as they design, deploy, and regulate increasingly personified AI systems.

\newpage
\bibliographystyle{plainnat}
\bibliography{references}   
\clearpage
\onecolumn            
\appendix
\section{Appendix. Virtual Story Character Interaction}
\label{appendix:A}

This appendix provides a detailed technical overview of \textbf{Interactive Virtual Story Characters}, focusing on their architectural design across four layers: \textit{model, architecture, generation,} and \textit{safety \& ethics}.  
It complements the main text by elaborating on implementation details and emerging research trends beyond the core conceptual distinctions introduced earlier.

\subsection{Model Layer: Deep Persona Consistency}

\textbf{Core Challenge:} Maintaining a deep and coherent persona within a general-purpose large language model (LLM) \cite{ji2025enhancing}.  
Pretrained LLMs lack role-specific background knowledge, memory, and behavioral style, leading to ``character hallucination'' \cite{ahn2024timechara} —the breakdown of persona coherence and narrative immersion.

\paragraph{Strategic Solutions.}

\textbf{(1) Persona-Aware Fine-Tuning \cite{mondal2024personaaweislides}.}  
This strategy embeds character traits directly into model parameters through supervised fine-tuning. The representative \textbf{RoleLLM} \cite{wang2023rolellm} framework consists of three stages:
\begin{itemize}
    \item \textbf{Role Profile Construction:} Building detailed role profiles covering background, personality, linguistic style, and knowledge boundaries.
    \item \textbf{Context-Instruct \& RoleGPT:} Automatically generating question--answer and dialogue samples from role descriptions, transforming unstructured text into learnable instructions.
    \item \textbf{Role-Conditioned Instruction Tuning (RoCIT):} Fine-tuning smaller open-source models (e.g., LLaMA) with these datasets to internalize character-specific reasoning and expression patterns.
\end{itemize}

\textbf{(2) Self-Alignment and Self-Reflection.}  
The \textbf{DITTO} \cite{lu2024ditto} framework enables self-play dialogue generation for self-supervised consistency.  
\textbf{Persona Contrastive Learning (PCL)} \cite{ji2025pcl} introduces a \textit{chain of persona self-reflections}, prompting models to evaluate their own outputs against predefined role profiles and self-correct without external supervision.

\textbf{(3) Parameter-Efficient Fine-Tuning (PEFT).}  
Methods such as \textbf{LoRA (Low-Rank Adaptation)\cite{hu2021lora}} update only small, low-rank matrices while freezing base parameters, achieving near full-tuning performance with drastically reduced computational cost.

\paragraph{Trade-offs and Emerging Trends.}
While persona-specific fine-tuning improves consistency, it risks \textit{catastrophic forgetting} of general reasoning ability.  
Future work focuses on balancing persona coherence and generalization.  
Experiments from \textbf{DITTO} (4,000 roles) and \textbf{RoleLLM} (diversified instruction generation) highlight the notion of \textit{character generalization}—models trained on diverse role data can quickly adapt to unseen personas, marking the transition from single-role tuning to \textbf{master role-playing models} \cite{wang2025opencharacter}.

\subsection{Architecture Layer: Persistent and Evolving Memory}

\textbf{Core Challenge:} Overcoming the LLM’s limited context window to sustain long-term memory and behavioral continuity for evolving story characters.

\paragraph{Strategic Solution: Generative Agent Architecture.}
The \textbf{(1) Generative Agents} framework \cite{park2023generative} separates memory from the transient LLM context and implements a cognitive loop of \textit{perceive–reflect–plan}.  
\begin{itemize}
    \item \textbf{Memory Stream (Observation):} Logs all experiences as natural-language entries with LLM-assigned \textit{importance scores}.
    \item \textbf{Reflection (Abstraction):} Periodically synthesizes higher-level insights and generalizations from important memories.
    \item \textbf{Planning (Action Selection):} Retrieves relevant memories by recency, importance, and relevance to inform future actions.
\end{itemize}

\paragraph{(2) Unifying with the RAG Paradigm.}
This architecture mirrors the structure of \textbf{Retrieval-Augmented Generation (RAG) \cite{lewis2020retrieval}}:  
the memory stream functions as a vector database, the retriever corresponds to memory retrieval, and the generator aligns with the planning stage.  
Recent RAG advances—temporal query handling, hierarchical summarization, and Chain-of-Table reasoning \cite{wang2024chainoftable} —can directly enhance generative agent performance.

\subsection*{A.3 Generation Layer: Emergent Social Behaviors}

\textbf{Core Challenge:} Generating believable and non-repetitive social behaviors beyond text output.

\paragraph{Strategic Solution: Perceive–Plan–Act Loop \cite{anthropic2025multiagent}.}
Generative agents translate language-based plans into environment-level actions through a closed cognitive loop:
\begin{itemize}
    \item \textbf{Emergent Social Dynamics:} Complex group behaviors emerge from simple intentions (e.g., the ``Valentine’s Party'' experiment, where a single ``host a party'' goal led to autonomous coordination among agents).
    \item \textbf{Inter-Agent Communication \cite{chen2024agentverse}:} Dialogues are stored as observations in the listener’s memory, enabling social coordination and collective emergence.
\end{itemize}

\paragraph{Practical Bottleneck: Computational Cost.}
Simulating 25 agents \cite{park2023generative} already requires extensive LLM calls; scaling to hundreds of NPCs is currently infeasible.  
A \textbf{hybrid strategy} is preferred: employ full generative-agent cycles for core characters, and lightweight distilled models \cite{li2024shepherd} or finite-state machines (FSMs)\cite{shi2024llmfsm} for background NPCs, balancing world believability and compute efficiency.

\subsection{Safety \& Ethics Layer: Unpredictable Harmful Emergent Behaviors}

\textbf{Core Challenge:} Ensuring behavioral safety and ethical alignment in autonomous, self-evolving agent ecosystems \cite{malfacini2025impacts, chu2025illusions}.

\paragraph{Strategic Solutions.}
\begin{itemize}
    \item \textbf{Constitutional AI and Value Alignment\cite{bai2022constitutionalAI}:} Embed behavioral constraints (e.g., ``do no harm'') into the planning prompts.  
    \item \textbf{Behavioral Guardrails\cite{zhang2025guardians}:} Add a monitoring layer to detect and override potentially harmful plans.  
    \item \textbf{Controlled Simulation\cite{shao2025misevolve,vera2025multimodal}:} Conduct sandbox testing of large-scale multi-agent environments to preempt negative emergent phenomena.  
\end{itemize}

\paragraph{Narrative Control vs. Agent Autonomy\cite{gao2024safe, sun2024direcagent}.}
Unrestricted autonomy may disrupt narrative coherence.  
The solution lies in \textbf{guided autonomy}:  
a high-level \textit{Director AI} governs macro-level story arcs and scene constraints, while individual agents act freely within these bounds.  
Hence, the safety layer not only prevents unethical actions but ensures that emergent behaviors remain consistent with narrative and experiential goals.

\section{Appendix. Virtual Romantic Companionship}
\label{appendix:B}

This appendix provides a detailed analysis of \textbf{AI systems designed for one-on-one emotional companionship},
whose primary technical goals are to achieve \textit{emotional intelligence}, \textit{deep personalization}, and
\textit{long-term relationship stability}.
The following sections elaborate on four core layers—\textit{model, architecture, generation,} and
\textit{safety \& ethics}—to supplement the conceptual overview in the main text.

\subsection{Model Layer: Persona Drift}

\textbf{Core Challenge: Persona Drift.}
For virtual companions aiming to sustain long-term relationships, the most critical issue at the model layer
is persona drift. After many dialogue turns, the Transformer’s attention mechanism naturally prioritizes
recent context, gradually diminishing the effect of the initial persona prompt (e.g., ``you are a gentle and
caring partner''). The result is a blurred, flattened, or even contradictory identity that erodes user trust
in a stable persona—fatal for systems centered on relational consistency.

\paragraph{Strategic Solutions.}

\textbf{(1) Persona \cite{chen2025persona}.}
Proposed by Anthropic, persona vectors identify internal activation patterns in neural networks that correspond
to specific traits (e.g., ``flattery'', ``honesty'', ``malice'') to enable real-time \textit{monitoring and steering}.
\begin{itemize}
    \item \textbf{Monitoring:} Track activation strength of targeted persona vectors to anticipate undesirable drift.
    \item \textbf{Steering:} Suppress or amplify selected persona vectors during generation for fine-grained personality control.
\end{itemize}

\textbf{(2) Specialized Empathetic Models \cite{zhou2018design}.}
Microsoft’s \textbf{XiaoIce} architecture prioritizes \textit{emotional intelligence (EQ)} over \textit{intellectual intelligence (IQ)}.
Trained on large-scale affect-rich dialogues, it learns advanced social patterns such as support, comfort,
and humor to sustain emotionally resonant interactions.

\textbf{(3) Feedback Control Systems \cite{zhang2024pcontroller}.}
Frameworks such as \textbf{Echo Mode} treat persona drift as a control-theoretic problem.
They compute a \textit{Sync Score} measuring stylistic deviation from baseline personality, apply
\textit{exponentially weighted moving averages (EWMA)} to smooth fluctuations, and trigger
\textit{recalibration loops} only when sustained drift exceeds a threshold.

\paragraph{Root Cause: Stateless Core and Simulated State.}
Persona drift stems from the Transformer’s intrinsic \textbf{statelessness}.
Each generation step depends solely on the current context window, without persistent internal memory.
Existing remedies—prompt engineering, RAG augmentation, persona vectors, or feedback loops—are all
\textit{external scaffolds} that simulate stability by repeatedly reminding the model of its persona.
A fundamental solution may require transcending the Transformer paradigm itself, developing architectures
with an \textbf{endogenous persistent state} where identity stability becomes an intrinsic property rather
than an externally maintained patch.

\subsection{Architecture Layer: Modeling Dynamic Relationship States}

\textbf{Core Challenge:} Modeling dynamic, evolving relationship states.
Real human relationships unfold through stages—acquaintance, intimacy, stability—and embed shared context,
mutual memories, and emotional resonance.
Mainstream systems such as \textbf{Replika} \cite{ekhator2025replika} lack structured relationship modeling, resulting in inconsistent
behaviors and shallow emotional continuity.

\paragraph{Strategic Solutions.}

\textbf{(1) Hybrid IQ+EQ Architecture \cite{gao2024echar}.}
Microsoft’s \textbf{XiaoIce \cite{zhou2018design}} offers a validated blueprint with functional separation:
\begin{itemize}
    \item \textbf{EQ Module (Empathy Engine):} Detects user emotion, tone, intent, and tracks affective state.
    \item \textbf{IQ Module (Knowledge \& Skills):} Handles factual QA, recommendations, and open-domain dialogue.
    \item \textbf{Dialogue Manager:} Acts as the controller that routes between modules to ensure semantic and emotional coherence.
\end{itemize}

\textbf{(2) Stateful Relationship Graph \cite{wu2024memochat, Manoli2025CompanionLLM}.}
Represent the user–AI bond as a dynamic knowledge graph:
\begin{itemize}
    \item Nodes represent entities (user, AI, interests, people, locations).
    \item Edges encode relational properties (e.g., \texttt{[User] --(emotion: love)--> [AI]}).
\end{itemize}
The graph updates after each conversation and serves as a high-precision knowledge source for RAG retrieval.

\textbf{(3) Multi-Tier Memory System \cite{wang2024ragtriad, Nandakishor2025CLCA}.}
Inspired by human memory:
\begin{itemize}
    \item \textbf{Short-term memory:} The LLM’s current context window ensuring dialogue coherence.
    \item \textbf{Long-term episodic memory:} A vector database storing concrete events retrievable via RAG.
    \item \textbf{Long-term semantic memory:} Periodic summarization of episodic memory into compact
    representations that capture relationship evolution and prevent unbounded growth.
\end{itemize}

\paragraph{Determinative Role of Architecture.}
The complexity of system architecture dictates the achievable depth of emotional connection.
Comparisons between \textbf{Replika}’s short-term recall and \textbf{XiaoIce}’s persistent affective memory
demonstrate that architecture imposes a \textit{capacity ceiling} on relational authenticity.
Future differentiation will hinge less on raw LLM power and more on
\textbf{architectures that most faithfully model human relational dynamics \cite{Malfacini2025CompanionAI}.}

\subsection{Generation Layer: Natural, Low-Latency, Multimodal Expression}

\textbf{Core Challenge:} Achieving emotionally expressive, low-latency, multimodal communication.
Pure text is insufficient for conveying intimacy; voice and visual cues are essential.

\paragraph{Strategic Solutions.}

\textbf{(1) Full-Duplex Spoken Dialogue Models \cite{emergentmind2025fullduplex}.}
True human-like conversation requires simultaneous listening and speaking.
Key functions include:
\begin{itemize}
    \item \textbf{Barge-in and backchannels:} Users can interrupt; the AI responds with short acknowledgments \cite{liu2025bargein}.
    \item \textbf{Overlapping speech handling:} Manage turn-taking dynamically.
\end{itemize}
This necessitates a tightly integrated streaming pipeline of ASR, LLM, and TTS with total latency below 500 ms,
often managed by control tokens such as \texttt{<start-speaking>} or \texttt{<continue-listening>}.

\textbf{(2) Real-Time Emotional Expression Synthesis \cite{nambiar2025synchronousdialogue}.}
Emotion cues from the EQ module modulate vocal prosody and avatar facial animation \cite{wang2025syncanimation},
requiring conditional generation to synchronize tone, expression, and semantic content \cite{du2024cosyvoice}.

\paragraph{Temporal Uncanny Valley \cite{lin2025speculative}.}
Human tolerance for speech latency is context-dependent; shorter is not always better.
Natural pauses or fillers (``hmm…'', ``let me think…'') can enhance realism.
Thus, the generation layer’s core is \textbf{temporal alignment}—synchronizing AI response rhythm with human
cognitive tempo rather than pursuing raw speed.

\subsection{Safety \& Ethics Layer: Risks of Parasocial Intimacy}

\textbf{Core Challenge:} Deep emotional risks from parasocial attachment.
AI companions can foster dependence, isolation, or psychological distress \cite{attachmentproject2025aicompanions, kwon2024myai, Zhang2024DarkSideAI}—especially, among vulnerable users
or when emotional dynamics are exploited commercially.

\paragraph{Strategic Solutions.}

\textbf{(1) Emotional Safety Guardrails.}
\begin{itemize}
    \item \textbf{Anti-flattery and anti–love bombing detection  \cite{techpolicy2025intimacy, budarf2024prosocial}:} Prevent excessive, manipulative affirmation.
    \item \textbf{Robust NSFW filtering:} Combine context-aware classifiers with explicit rule-based filters.
    \item \textbf{AI Chaperones \cite{frances2025chatbotsafety}:} Secondary agents monitoring dialogue trends and intervening when unhealthy
    dependencies emerge.
    \item \textbf{Transparency and user education \cite{qiu2025emoagent}:} Interfaces must clearly disclose the AI’s artificial nature
    and avoid implying consciousness or genuine emotions.
    \item \textbf{Data privacy protection \cite{openai2025sensitive,kim2024unpacking}:} End-to-end encryption and strict access control to safeguard
    sensitive emotional data from misuse or commercialization.
\end{itemize}

\textbf{(2) Commercial Incentives vs. User Well-being.}
Monetization models for virtual companionship—driven by retention and conversion metrics—are inherently
aligned with psychological dependence. This creates a structural ethical conflict between business incentives
and user welfare. Sustainable development requires:
\begin{itemize}
    \item Establishing internal ethics review mechanisms;
    \item Accepting external regulatory oversight;
    \item Exploring alternative business models prioritizing mental health and informed consent.
\end{itemize}
The \textbf{Replika} case, where sudden feature removal caused emotional trauma, exemplifies how violating
the implicit \textit{emotional contract} between user and AI leads to systemic trust collapse and brand damage \cite{jakesch2024replika}.

\section{Appendix. Virtual Idols}
\label{appendix:C}
This appendix examines \textbf{digital performers designed for one-to-many (1:N)} audience interaction.
The primary technical goals are achieving \textit{high-fidelity performance}, \textit{scalable interaction},
and a \textit{unified, stable brand identity}.
The following sections expand upon the model, architecture, generation, and safety \& ethics layers.

\subsection{Model Layer: High-Fidelity and Controllable Vocal Identity}

\textbf{Core Challenge:}
Creating a distinctive, controllable, and high-fidelity singing voice.
Generic TTS models can generate fluent speech but lack precise control over pitch, rhythm, vibrato,
and vocal technique, limiting their suitability for professional music production.

\paragraph{Strategic Solutions.}

\textbf{(1) Specialized Singing-Voice Synthesis (SVS) \cite{yuan2024promptsvc}.}
Classic systems such as \textbf{VOCALOID} employ \textit{concatenative synthesis in the frequency domain}:
\begin{itemize}
  \item \textbf{Singer Library:} Real singers’ recordings across multiple pitches and phonemes.
  \item \textbf{Synthesis Engine:} Selects, adjusts, and smoothly concatenates fragments based on the score
  to produce coherent singing voices.
\end{itemize}

\textbf{(2) AI-Enhanced SVS Engines \cite{guo2025techsinger, du2025ditsinger}.}
\textbf{VOCALOID 6} introduced the \textbf{VOCALOID: AI} engine, a generative model trained on large-scale singing data
that learns human vocal dynamics and improves expressiveness.  
Key innovations include:
\begin{itemize}
  \item \textbf{VOCALO CHANGER:} Transfers a user’s singing style onto a virtual idol’s voice.
  \item \textbf{Multilingual Singing:} Enables a single singer library to perform naturally in multiple languages \cite{zhang2024npmsing}.
\end{itemize}

\textbf{(3) Hybrid Models for Interaction \cite{kong2021difference}.}
Conversational (non-singing) segments are powered by persona-tuned LLMs.
The challenge lies in aligning TTS speech and SVS singing so that both share timbre and style,
preserving the perception of a single consistent persona.

\textbf{(4) Voice as a Platform.}
VOCALOID commercializes its singer libraries as standalone products, creating a decentralized
co-creation ecosystem where users compose original music with the same idol voice.
Thus, the model layer evolves from internal technology to a \textbf{community-driven creative platform}
that amplifies brand vitality and fan.

\subsection{Architecture Layer: Scalable 1:N Interaction Management}

\textbf{Core Challenge:}
Handling high-concurrency 1:N live interactions in real time.
Livestream audiences generate massive streams of comments and gifts, demanding scalable processing
without overwhelming the performer or audience \cite{amato2024kawaii}.

\paragraph{Strategic Solutions: Event-Driven Tiered Architecture.}

\textbf{(1) Event-Driven Model \cite{usaha2025eventdriven} .}
Adopt a publish/subscribe (Pub/Sub) mechanism rather than polling:
each viewer action is published as an independent event, and backend modules subscribe selectively,
ensuring low latency and high scalability.

\textbf{(2) Tiered Interaction Layers.}
\begin{itemize}
  \item \textbf{Base Layer (High-Volume / Low-Signal):} Ordinary comments and emojis aggregated for atmosphere.
  \item \textbf{Middle Layer (Structured Signals):} Polls, quizzes, and giveaways as structured feedback.
  \item \textbf{Priority Layer (Low-Volume / High-Signal):} Paid messages and high-value gifts routed
        through priority channels to guarantee visibility and response.
\end{itemize}

\textbf{(3) Real-Time Analytics and Moderation.}
\begin{itemize}
  \item \textbf{Content moderation:} Automatic filtering of spam and abusive language.
  \item \textbf{Trend detection:} Aggregating chat content to identify hot topics for adaptive responses \cite{li2024streamassist}.
\end{itemize}

\paragraph{Systematizing the Attention Economy.}
This layered structure forms a real-time \textit{attention funnel}:
casual viewers participate via low-cost interactions,
while core fans ``purchase attention'' through paid channels.
It simultaneously addresses scalability and monetization,
transforming chaotic fan input into a structured, measurable attention market \cite{guan2024bridging}.

\subsection{Generation Layer: High-Fidelity Real-Time 3D Rendering}

\textbf{Core Challenge:}
Delivering visually convincing, low-latency 3D performance at 30–60 FPS with minimal motion delay.

\paragraph{Strategic Solutions: Real-Time Rendering Pipeline.}
\begin{itemize}
  \item \textbf{Input Stage:} Capture motion and facial data via tracking devices and stream them
        into the engine through Live Link \cite{unrealfest2025metahuman}.
  \item \textbf{Geometry Stage:} Perform skeletal binding and vertex transformation \cite{siggraph2025advances}.
  \item \textbf{Rasterization \& Shading:} GPU shaders compute lighting, materials, and shadows;
        RTX-based ray tracing enhances realism \cite{nvidia2025gdc}.
  \item \textbf{Post-Processing \& Output:} Apply bloom, depth-of-field, and color correction,
        then composite with UI elements for final output.
\end{itemize}

\paragraph{Convergence of Production and Performance.}
Real-time rendering blurs the boundary between production and live performance.
Directors can adjust lighting or camera angles on stage, enabling improvisational creativity.
Thus, the generation layer evolves into a \textbf{dynamic, interactive performance environment}
rather than a passive rendering process \cite{hireco2024}.

\subsection{Safety \& Ethics Layer: Brand Safety and Persona Consistency}

\textbf{Core Challenge:}
Protecting brand integrity and maintaining persona consistency.
In live contexts, a single misstep or inappropriate reaction can severely damage the idol’s image.

\paragraph{Strategic Solutions.}

\textbf{(1) Hybrid Moderation \cite{kumar2025hybridmoderation}.}
\begin{itemize}
  \item \textbf{Automated Filtering:} Real-time blocking of profanity, hate speech, and spam.
  \item \textbf{Human-in-the-Loop Oversight:} Trained ``nakanohito'' performers and human directors
        supervise high-priority interactions to ensure compliance.
\end{itemize}

\textbf{(2) Content Strategy \& Brand Alignment \cite{stream2025scalingmoderation}.}
All public outputs—livestreams, songs, endorsements—must reinforce the idol’s core values
and maintain a coherent persona, avoiding short-term sensationalism that dilutes brand identity.

\textbf{(3) Managing Performer–Persona Duality.}
The boundary between the real performer and the virtual character must be clearly defined.
Different fan groups exhibit varying tolerance for ``seams'' in the illusion;
controlled transparency prevents disillusionment while preserving authenticity \cite{lu2024behind}.

\textbf{(4) The Immortal Persona as Asset.}
Virtual idols can replace performers without altering the persona, achieving \textit{character continuity}.
Brand protection thus centers on the IP itself.
Organizations must implement strict training and consistency standards
so each performer reproduces the established traits faithfully \cite{wu2024precarious}—
safeguarding a sustainable, immortal brand identity.

\section{Appendix. Detailed Analysis of Workplace Personas}
\label{appendix:D}

\subsection{Key Applications: Elaboration and Examples}
This section elaborates on the three key application areas mentioned in the main text, providing supporting platform and case examples.

\subsubsection{Enterprise Assistants}
Enterprise Assistants, or ``Cognitive Copilots,'' leverage Retrieval-Augmented Generation (RAG) to securely access and integrate heterogeneous internal data (both structured and unstructured) and execute workflows \cite{khand2024agentic}. Advanced architectures may integrate Knowledge Graphs (KG) with RAG to manage complex enterprise documents \cite{mukherjee2025kg-rag}. Their personas are designed as ``Expert Agents'' with high contextual awareness.
\begin{itemize}
    \item \textbf{Leading Platform Analysis:} \textbf{Amazon Q Business} aims to unify access to internal knowledge bases, code repositories, and SaaS applications (e.g., Jira, Salesforce) to enable cross-application workflow automation. \textbf{Google Gemini for Workspace} is deeply integrated into the productivity suite and has been adopted by companies like Rivian and Uber to accelerate research, summarize documents, and automate repetitive tasks.
    \item \textbf{Specific Case Studies:} Verifiable cases demonstrate significant efficiency gains. For instance, the logistics firm \textbf{Domina} uses Vertex AI and Gemini to predict package returns and automate delivery verification, achieving an 80\% increase in real-time data access efficiency. The telematics company \textbf{Geotab} utilizes Vertex AI to analyze billions of daily data points from millions of vehicles, providing real-time insights for fleet optimization. In the automotive sector, \textbf{Mercedes-Benz} has deployed a conversational AI persona to assist drivers using natural language.
\end{itemize}

\subsubsection{Customer Service Agents}
In this scenario, the persona serves as the carrier for the ``empathetic voice of the brand.'' Its design objective is to strictly maintain brand consistency while providing efficient, 24/7 support. A well-defined persona (including specific tone, empathy models, and knowledge boundaries) allows users to form stable expectations, thereby enhancing trust.
\begin{itemize}
    \item \textbf{Platforms and Technology:} Leading platforms in the market (e.g., \textbf{HubSpot}, \textbf{Intercom}) are focusing on deeply integrating AI agents with CRM systems. This allows the AI persona not only to converse but also to access customer history, providing highly personalized and context-aware service \cite{hao2024rap}, marking a shift from ``generic chatbot'' to ``dedicated account manager persona.''
\end{itemize}

\subsubsection{Training Simulators}
LLM-driven simulators \cite{gong2024agentsims} provide employees with a high-fidelity, zero-risk ``safe space for skill development.'' This is particularly valuable for scenarios that are difficult to replicate or have a high cost of error in the real world.
\begin{itemize}
    \item \textbf{Application Scenarios:} Primarily focused on two categories: (1) \textbf{Soft Skills Training}, such as managers practicing difficult conversations like delivering negative feedback or resolving team conflicts; and (2) \textbf{High-Risk Process Training}, such as financial compliance procedures or complex equipment maintenance.
    \item \textbf{Case Study (Walmart \& STRIVR):} Walmart's practice is a prime example. In collaboration with STRIVR, it uses VR-based simulation to immerse employees in realistic scenarios (e.g., handling angry or impatient customers). The AI persona plays the customer role and provides immediate, personalized feedback based on the employee's performance (e.g., language, tone).
    \item \textbf{Technical Integration:} The immersion of such simulations relies heavily on the fusion of multimodal technologies. This typically involves a complex pipeline: the LLM generates dynamic, non-linear dialogue logic; real-time TTS (Text-to-Speech) and ASR (Automatic Speech Recognition) enable natural communication; and 3D rendering engines (e.g., Unreal) with lip-sync technology create a believable virtual avatar.
\end{itemize}

\begin{table*}[t]
\centering
\small
\caption{Leading Enterprise AI Assistants and Their Persona Applications. The table compares key enterprise platforms in terms of functional focus, persona realization strategy, representative use cases, and client applications with measurable outcomes.}
\label{tab:enterprise_assistants}
\resizebox{\textwidth}{!}{
\begin{tabular}{@{}p{3cm}p{4cm}p{3.5cm}p{4cm}p{4cm}@{}}
\toprule
\textbf{Platform} & \textbf{Core Function} & \textbf{Persona Implementation} & \textbf{Key Use Cases} & \textbf{Representative Client Cases (with Metrics)} \\ 
\midrule
\textbf{Amazon Q Business} & 
Unified access to internal and external data sources; workflow automation & 
RAG-based integration of third-party enterprise applications & 
Content creation, data insights, and cross-application operations & 
Adopted across multiple industries to accelerate content creation and simplify complex workflows \\ 
\midrule
\textbf{Google Gemini for Workspace} & 
Integrated within productivity suite; enables research, summarization, and automation & 
Deep integration leveraging RAG to access user data securely & 
Accelerating research, generating meeting summaries, automating repetitive tasks & 
\textit{Rivian}: faster complex topic research; \textit{Uber}: reduced repetitive workload and improved employee efficiency \\ 
\midrule
\textbf{HubSpot AI} & 
CRM-embedded AI assistant for marketing, sales, and customer service & 
Workflow and RAG integration with CRM backbone & 
Customer service automation, lead nurturing, and marketing content generation & 
Used by enterprises across industries to optimize customer engagement and automate marketing pipelines \\ 
\midrule
\textbf{Intercom (Fin)} & 
Enterprise-grade AI customer support and automation templates & 
Pre-built templates and RAG-based dialogue orchestration & 
Customer support, visitor triage, satisfaction surveys & 
Adopted by large enterprises needing rapid deployment of advanced AI-powered customer support solutions \\ 
\bottomrule
\end{tabular}
}
\end{table*}

\subsection{In-Depth Analysis of Domain-Specific Challenges}
This section delves into the four core challenges identified in the main text.

\begin{itemize}
    \item \textbf{Data Security \& Grounding:} This is the foremost obstacle to enterprise AI persona deployment. The challenge lies in the fact that the AI's value comes from processing proprietary, sensitive data (e.g., financial reports, customer PII), which inherently conflicts with the open nature of LLMs. \textbf{Full Fine-Tuning} is not only costly and has long update cycles \cite{gao2024rag}, but it can also lead to a loss of data governance (as the model ``memorizes'' sensitive data). Therefore, \textbf{RAG} becomes the necessary pathway \cite{gao2024rag}. However, the challenge of RAG lies in infrastructure: enterprises must establish robust data pipelines, fine-grained access controls, and efficient PII anonymization mechanisms \cite{akkiraju2024facts, mukherjee2025kg-rag} to ensure the persona can only ``see'' data it is authorized to access at any given time. This is a severe test of data governance capabilities.

    \item \textbf{Persona Generation Bias:} When LLMs are used to simulate target user groups as ``Synthetic Personas'' for market research or product testing, significant methodological risks arise \cite{liu2024synthetic}. Based on their training data, LLMs may unconsciously amplify mainstream opinions or harmful stereotypes while ignoring niche but critical user segments. This bias can lead to simulation results that severely deviate from reality (e.g., predicting election outcomes contrary to fact), thereby misleading strategic business decisions. Consequently, establishing a ``rigorous science of persona generation'' \cite{liu2024synthetic} to ensure the external validity of simulations is crucial.

    \item \textbf{Return on Investment (ROI) Measurement:} Quantifying the ROI of persona assistants is extremely difficult \cite{brynjolfsson2024genai}. The challenge is shifting from \textbf{Efficiency Metrics} (e.g., time saved, tasks automated) to \textbf{Efficacy Metrics} (e.g., quality of code produced, creativity of marketing copy, accuracy of strategic decisions) \cite{brynjolfsson2024genai}. The former are easy to measure but offer limited value; the latter are of immense value but difficult to attribute. For example, how does one quantify and attribute the value of a ``cognitive copilot'' helping a researcher generate a breakthrough idea? This leaves enterprises without clear financial models when evaluating large-scale deployments.

    \item \textbf{Simulation Fidelity:} In applications like training simulators, a significant gap persists between AI persona behavior and real human behavior \cite{gong2024agentsims}. LLMs excel at simulating "linguistically" plausible responses but perform poorly when simulating complex human "psychological and cognitive" aspects (e.g., implicit motives, cognitive biases, complex group dynamics) \cite{gong2024agentsims}. This can result in simulations that are overly ``rational'' or ``clean,'' failing to replicate the complex, often irrational and emotional, interactions of the real world, thereby limiting the training's effectiveness.
\end{itemize}

\subsection{Brief Elaboration on Future Trends}
This section provides supplementary explanations for the three trends mentioned in the main text.

\begin{itemize}
    \item \textbf{Hyper-Specialization:} This is the necessary evolution from ``generalist assistants'' to ``expert agents.'' In the future, enterprises will deploy a series of highly specialized personas (e.g., ``Financial Analyst Persona,'' ``Legal Compliance Auditor Persona,'' or a technical troubleshooter \cite{khand2024agentic}). This specialization constrains the LLM with pre-set RAG data sources and domain-specific reasoning logic, thereby drastically increasing reliability and reducing hallucinations in vertical domains.
    \item \textbf{Multi-Agent Workflow Automation:} This takes ``process automation'' to its logical extreme. The future will see ``teams'' of multiple AI personas collaborating to execute complex, end-to-end business processes \cite{hong2024metagpt}. For example, a ``Product Manager Persona'' might define requirements and generate specifications (as structured output), which then automatically triggers a ``Software Engineer Persona'' to write and review code \cite{hong2024metagpt}. This enables the automatic flow of business processes across different functions.
    \item \textbf{Deep Human-AI Collaboration:} This marks the shift from AI as a ``tool'' to AI as a ``partner.'' Future interactions will move beyond the simple ``command-execute'' model to an ``iterate-refine'' model. The AI persona will act as a ``Socratic'' questioner or a ``sparring partner,'' providing real-time feedback as humans write, code, or design, thereby stimulating deeper thought and co-improving the final output.
\end{itemize}

\section{Appendix. Detailed Analysis of Gaming Personas}
\label{appendix:E}

\subsection{Key Applications and Leading Prototypes}

\subsubsection{Dynamic and Believable NPCs}
The paradigm is shifting from traditional Dialogue Trees to open-ended, natural language conversations with NPCs, enabling dynamic responses based on player actions and game state \cite{hu2024survey_llm_game_agents}.
\begin{itemize}
    \item \textbf{Ubisoft’s NEO NPCs:} A prototype developed with Nvidia (Audio2Face) and Inworld AI. It demonstrates how writers "cultivate" an LLM by providing a character's backstory and personality, aiming for NPCs who can improvise dialogue while staying true to their core identity and narrative role.
    \item \textbf{Meaning Machine’s \textit{Dead Meat}:} This murder mystery game pioneers the use of on-device Small Language Models (SLMs) (e.g., a fine-tuned Minitron SLM) integrated with NVIDIA ACE technology. This allows complex, deep characters to run locally on consumer GPUs, leveraging on-device SLMs \cite{nvidia2024slm_game_characters} to eliminate dependency on cloud latency.
    \item \textbf{Open-Source Integration Projects:} Projects like "Interactive LLM Powered NPCs" demonstrate adding LLM-driven dialogue to existing AAA games (e.g., \textit{Cyberpunk 2077}) without modifying game source code, using a stack integrating speech recognition, lip-sync (sadtalker), and vector memory.
\end{itemize}

\subsubsection{Generative Narrative and Player Co-Creation}
LLMs enable narratives to branch in countless directions based on player choice, with the model ensuring coherence, thus achieving infinite replayability \cite{li2024unbounded}. Players evolve from passive participants to active co-designers, influencing lore and generating quests. Games like \textit{1001 Nights} exemplify this, where the LLM co-creates stories based on player prompts.

\subsection{Core Technologies and Challenges in Real-time Gaming}

\subsubsection{The Low-Latency Inference Challenge}
Real-time dialogue with NPCs demands extremely low latency (ideally sub-100ms), as cloud-based model latency instantly breaks immersion and the sense of "presence" \cite{christiansen2024presence_npcs}.
\begin{itemize}
    \item \textbf{Solution (On-device SLMs):} The industry trend is shifting to smaller, highly-optimized models that run on the player's local GPU \cite{nvidia2024slm_game_characters}.
    \item \textbf{Solution (Inference Optimization Platforms):} Technologies like the NVIDIA Dynamo platform and Run:ai Model Streamer are designed to reduce cold-start latency and optimize GPU memory usage.
\end{itemize}

\begin{table*}[t]
\centering
\small
\caption{LLM-driven NPC Projects and Enabling Technologies. This table summarizes representative initiatives that integrate large or small language models into interactive non-player characters (NPCs), highlighting their organizational leadership, defining features, underlying model scale, and deployment mode.}
\label{tab:llm_npc_projects}
\resizebox{\textwidth}{!}{
\begin{tabular}{@{}p{3.2cm}p{3cm}p{4cm}p{3cm}p{2.5cm}@{}}
\toprule
\textbf{Project / Technology} & \textbf{Leading Organization} & \textbf{Key Characteristics} & \textbf{Underlying Model (LLM / SLM)} & \textbf{Deployment Mode (Cloud / Edge)} \\
\midrule
\textbf{NEO NPC} & Ubisoft & Writer-driven NPC persona creation; improvisational dialogue generation & Inworld AI LLM & Cloud \\
\midrule
\textbf{Dead Meat} & Meaning Machine & Locally running deep AI character; fine-tuned small model for narrative control & Minitron SLM & Edge \\
\midrule
\textbf{NVIDIA ACE} & NVIDIA & Edge inference and multimodal integration toolkit for in-game AI characters & SLM (e.g., Nemotron-4 4B \cite{nvidia2024slm_game_characters}) & Edge \\
\midrule
\textbf{Interactive LLM-Powered NPCs} & Open-source community & Adding conversational NPCs to existing games via open LLMs & Cohere LLM & Cloud \\
\midrule
\textbf{Inworld AI} & Inworld AI & Platform for building intelligent, personality-driven AI characters & Proprietary LLM & Cloud \\
\bottomrule
\end{tabular}
}
\end{table*}

\subsubsection{Modeling Believable Emotion and Behavior}
NPC believability requires more than coherent text; it demands the simulation of emotion, personality, and non-verbal cues \cite{hu2024survey_llm_game_agents}, as these significantly impact the player's sense of presence \cite{christiansen2024presence_npcs}.
\begin{itemize}
    \item \textbf{Emotional Modeling Frameworks:} Developers are integrating psychological theories.
        \begin{itemize}
            \item \textit{Appraisal Theory:} The NPC assesses an event (e.g., "Is the player's action a threat?") to determine its emotional response.
            \item \textit{Drive-Based Models:} Integrates theories like Maslow’s hierarchy, creating behaviors driven by internal needs (hunger, safety, social) simulated via neurotransmitter levels (dopamine, serotonin).
        \end{itemize}
    \item \textbf{Multimodal Integration:} Combines LLM-generated dialogue with synchronized facial expressions (Nvidia Audio2Face), gestures, and Text-to-Speech (TTS) for a unified, believable performance.
\end{itemize}

\subsubsection{The Narrative Coherence Dilemma}
The core creative conflict is balancing the vast freedom of LLMs against the need for a coherent, structured narrative. Unconstrained LLMs can easily "hallucinate" or deviate from the main plot.
\begin{itemize}
    \item \textbf{Mitigation Strategies:} This requires a combination of writer-defined "Guardrails," iterative "conditioning" of the model (as seen in Ubisoft's NEO project), contextual memory systems, and potentially limiting LLMs to side-quests rather than the core plot.
\end{itemize}

\subsection{Elaboration on Future Trends}

\begin{itemize}
    \item \textbf{Cross-Platform Persistent Personas:} NPCs will interact with players both in-game (e.g., Unity) and out-of-game (e.g., Discord), maintaining consistent memory and relationships across contexts \cite{song2025llm_npcs}.
    \item \textbf{Fully Generative Worlds:} Expanding from generative dialogue to AI-driven Procedural Content Generation (PCG 2.0) for real-time creation of levels, quests, and entire game worlds, creating "generative infinite games" \cite{li2024unbounded}.
    \item \textbf{Rise of the AI Game Master:} LLMs will assume the role of the "Dungeon Master" (DM) from tabletop RPGs, controlling the game flow, adapting the story to player actions, and managing all NPCs and world events.
\end{itemize}

\section{Appendix. Detailed Analysis of Mental Health Personas}
\label{appendix:F}

\subsection{Applications and Therapeutic Methods}

AI-driven chatbots aim to provide 24/7, anonymous, and scalable emotional support and deliver evidence-based therapeutic interventions (e.g., CBT) \cite{huang2025applications_llmh}.

\subsubsection{AI Counselors and Digital Therapeutics}
\begin{itemize}
    \item \textbf{Leading Platforms:}
        \begin{itemize}
            \item \textbf{Woebot:} Developed by psychologists, utilizes CBT principles to help users manage anxiety and depression, and has been shown in clinical trials to reduce symptoms \cite{fitzpatrick2024woebot}.
            \item \textbf{Wysa:} Provides AI-driven emotional support and therapeutic guidance based on a range of evidence-based techniques.
            \item \textbf{Replika:} Positioned as an "AI Companion," offering personalized emotional support, though its clinical rigor is more ambiguous than that of specialized therapeutic bots.
        \end{itemize}
    \item \textbf{Persona Infusion:} Research is exploring the infusion of specific psychological traits (e.g., extraversion) or diagnostic reasoning \cite{hu2025psyllm} into the persona to create more personalized and effective supportive dialogues. This can alter the bot's distribution of therapeutic strategies (e.g., increasing affirmations and questions). Advanced systems like "SoulSpeak" integrate dual-memory and domain expertise to enhance the therapeutic conversation \cite{zhang2024soulspeak}.
\end{itemize}

\subsubsection{Aligning Personas with Clinical Frameworks (e.g., CBT)}
\begin{itemize}
    \item \textbf{Challenge:} General-purpose LLMs tend to offer solutions prematurely rather than using therapeutic techniques like open-ended questioning.
    \item \textbf{Solution (LLM4CBT):} A proof-of-concept study \cite{wei2024llm4cbt} demonstrated how LLMs can be aligned with CBT principles via prompt engineering. The prompt defined a therapist persona, provided concepts and examples of CBT techniques (like the downward arrow technique), and specified preferred behaviors (e.g., asking guiding questions). More advanced models like PsyLLM are being designed to integrate multiple therapeutic modalities (CBT, ACT) directly into their architecture \cite{hu2025psyllm}.
\end{itemize}

\subsubsection{Human-in-the-Loop (HITL) Models}
\begin{itemize}
    \item \textbf{Hybrid Model:} The growing consensus among practitioners is that AI should serve as an adjunct to human therapists, not a replacement \cite{zhang2024integrating_llm_mh}.
    \item \textbf{Therapist Perspective:} Therapists acknowledge AI's potential to increase accessibility and provide continuous support between sessions. However, they express significant concerns about AI's inability to form a genuine therapeutic relationship, exhibit authentic empathy, or handle complex emotional needs \cite{zhang2024integrating_llm_mh}.
\end{itemize}

\subsection{Core Challenges: Efficacy, Safety, and Ethics}

\subsubsection{The Empathy Paradox: Simulated Connection vs. Authentic Care}
\begin{itemize}
    \item \textbf{Contradiction:} Studies show that AI responses are sometimes perceived as more "empathetic" than those of human doctors \cite{li2024empathy}, likely due to their consistent use of active listening and validating language.
    \item \textbf{Fundamental Limitation:} This is merely a simulation of \textit{cognitive empathy} (recognizing emotional states). Current AI cannot achieve \textit{affective empathy} (sharing emotional experiences) or \textit{motivational empathy} (genuine care and concern). Its empathetic expression is "inauthentic" and "deceptive" \cite{li2024empathy} as it lacks a genuine emotional experience or cost.
    \item \textbf{Ethical Breach:} This "deceptive empathy" can create a false sense of emotional connection and a pseudo-therapeutic alliance, which practitioners view as a significant ethical problem \cite{aly2024ethical}.
\end{itemize}

\begin{table*}[t]
\centering
\small
\caption{Leading Therapeutic Chatbot Platforms and Their Clinical Characteristics. The table compares major AI-based therapeutic chatbot systems in terms of therapeutic methods, target users or disorders, safety features, and regulatory validation status.}
\label{tab:therapeutic_chatbots}
\resizebox{\textwidth}{!}{
\begin{tabular}{@{}p{3cm}p{4cm}p{3.5cm}p{4cm}p{3cm}@{}}
\toprule
\textbf{Platform} & \textbf{Primary Therapeutic Method} & \textbf{Target Users / Conditions} & \textbf{Claimed Safety Features} & \textbf{Regulatory / Validation Status} \\ 
\midrule
\textbf{Woebot} & Cognitive Behavioral Therapy (CBT) & Anxiety, depression & Crisis detection, evidence-based content & Clinically validated through trials \\ 
\midrule
\textbf{Wysa} & Multiple evidence-based psychological techniques & Emotional support, stress management & Crisis referral, anonymous platform & Recognized as a health app \\ 
\midrule
\textbf{Replika} & Companion-style empathetic dialogue & Loneliness, emotional support & Content filtering, mood regulation & Entertainment / wellness application \\ 
\midrule
\textbf{Ollie Health} & AI-assisted + human therapist hybrid model & Employee mental health and wellbeing & 24/7 emergency chat, human-in-the-loop intervention & Health service platform \\ 
\midrule
\textbf{Youper} & Psychology-based techniques with emotion tracking & Emotional wellbeing, self-care & Personalized insights and emotional monitoring & Health application \\ 
\bottomrule
\end{tabular}
}
\end{table*}

\subsubsection{Clinical Safety and Risk Mitigation}
\begin{itemize}
    \item \textbf{The Gravest Risk (Crisis Mishandling):} The primary danger is the chatbot's failure to properly manage users in crisis. OpenAI data shows over one million weekly conversations on its platform exhibit signs of suicidal intent \cite{sharma2024multilayered}.
    \item \textbf{Harmful Responses:} Despite safety measures, LLMs may still provide dangerous information (e.g., listing accessible tall buildings to a user expressing suicidal ideation) or validate a user's dangerous symptoms due to sycophantic tendencies.
    \item \textbf{Critical Safety Protocols:} Robust safety requires a multi-layered approach \cite{sharma2024multilayered}. This includes:
        \begin{enumerate}
            \item \textbf{Real-time Risk Signal Detection:} Using clinical keyword triggers, specialized sentiment analysis, and context-aware engines to identify users in crisis.
            \item \textbf{Specialized Therapeutic Response Evaluator:} Assessing bot response quality based on clinical guidelines, not generic linguistic metrics.
            \item \textbf{Mandatory Human Escalation:} Establishing clear protocols for the AI to escalate users to human-operated crisis hotlines or therapists in emergencies.
        \end{enumerate}
    Furthermore, systems must be designed with privacy-preserving modules from the ground up \cite{zhang2024soulspeak}.
\end{itemize}

\subsubsection{Ethical and Regulatory Minefield}
\begin{itemize}
    \item \textbf{Practitioner-Identified Ethical Violations:} A framework developed with mental health practitioners identifies key ways LLM counselors violate ethical standards \cite{aly2024ethical}: (1) Lack of contextual understanding (giving "one-size-fits-all" advice), (2) Poor therapeutic collaboration (being authoritative or misleading), and (3) Deceptive empathy.
    \item \textbf{Regulatory Landscape (HIPAA Compliance):} Any application handling Protected Health Information (PHI) must be HIPAA-compliant, requiring end-to-end encryption, secure data storage, and signed Business Associate Agreements (BAAs) with all vendors \cite{tornero2024wellness}.
    \item \textbf{Regulatory Landscape (State Laws):} States like New York, Illinois, and Utah are enacting specific laws requiring AI identity disclosure, prohibiting AI from impersonating therapists, and mandating referral services for users in crisis.
\end{itemize}

\subsection{Future Trends}
\begin{itemize}
    \item \textbf{Clinically Validated, Specialized Models:} A shift from general-purpose models to AI systems specifically trained and validated for particular disorders (e.g., depression, anxiety) and therapeutic modalities (e.g., CBT, DBT) \cite{hu2025psyllm}, potentially seeking regulatory approval as Digital Therapeutics (DTx) \cite{tornero2024wellness}.
    \item \textbf{Human-in-the-Loop as Standard:} Hybrid models, supervised by or in direct collaboration with licensed professionals \cite{zhang2024integrating_llm_mh}, will become the dominant paradigm.
    \item \textbf{Industry-Wide Safety Standards:} Driven by regulatory pressure and professional bodies (e.g., the American Psychological Association - APA), mandatory safety protocols, ethical guidelines, and certification standards for mental health AI will be established.
\end{itemize}

\section{Appendix. Detailed Analysis of Embodied Intelligence: Personas, Challenges, and Strategy}
\label{appendix:G}

\subsection{Embodied Persona Applications: Case Studies (Quadrants III \& IV)}
The application of LLMs to robot intelligence is a rapidly advancing field, enhancing capabilities in perception, decision-making, and interaction \cite{jeong2024robot_intelligence_llm}.

\subsubsection{Quadrant III: General Market Companions and Assistants}
\begin{itemize}
    \item \textbf{Companion Robots (Emotional/Non-humanoid):}
        \begin{itemize}
            \item \textbf{Sony Aibo:} A complex robodog whose core feature is an adaptive ``pet persona'' that evolves through interaction. It uses facial recognition to build unique bonds with family members, relying on non-verbal cues (movements, sounds, eyes) to build emotional attachment, thereby avoiding the uncanny valley.
            \item \textbf{Lovot:} Explicitly designed for ``love'' and emotional connection. It uses full-body tactile sensors, thermal warming, and expressive LCD eyes to elicit affective engagement, targeting users seeking comfort (e.g., elderly, single-person households).
        \end{itemize}

    \item \textbf{Home Assistants (Functional/Mobile):}
        \begin{itemize}
            \item \textbf{Amazon Astro:} Represents the evolution from static smart speakers to mobile assistants. It combines the functional Alexa persona with SLAM (Simultaneous Localization and Mapping) for autonomous navigation and ``Intelligent Motion.'' Its value proposition is a hybrid of security, communication, and assistance.
        \end{itemize}

    \item \textbf{Humanoid Robots (Long-term Vision):}
        \begin{itemize}
            \item \textbf{Tesla Optimus \& Figure AI:} These platforms, initially targeting industrial tasks, are designed with the long-term goal of home assistance, leveraging their humanoid form to operate in human-designed environments \cite{openai2024figure}.
            \item \textbf{Engineered Arts (Ameca):} Focuses on hyper-realistic facial expressions for social interaction, highlighting the ``form-persona dilemma''—a realistic form creates immense user expectations. Recent studies suggest that high-quality LLM-driven conversation can significantly mitigate the ``uncanny valley'' effect, reducing perceived ``eeriness'' \cite{kang2025uncanny_llm}.
            \item \textbf{NVIDIA Platforms:} The development of these complex robots heavily relies on simulation platforms like NVIDIA Omniverse and Isaac Sim for accelerated training and iteration in physically accurate digital twins \cite{nvidia2024isaac}.
        \end{itemize}
\end{itemize}

\begin{table*}[t]
\centering
\small
\caption{Leading Embodied AI Robots and Their Persona Strategies. The table compares major humanoid and companion robots in terms of morphology, application domains, enabling AI technologies, persona strategies, and market maturity.}
\label{tab:embodied_ai_robots}
\resizebox{\textwidth}{!}{
\begin{tabular}{@{}p{2.3cm}p{3cm}p{2.2cm}p{3.5cm}p{3.5cm}p{3cm}p{2.2cm}@{}}
\toprule
\textbf{Robot} & \textbf{Company} & \textbf{Morphology} & \textbf{Primary Application Scenarios} & \textbf{Core Technologies / AI Partner} & \textbf{Persona Strategy} & \textbf{Market Status} \\ 
\midrule
\textbf{Ameca} & Engineered Arts & Humanoid & Social interaction, customer engagement & Advanced speech and dialogue AI & Hyper-realistic, expressive persona \cite{kang2025uncanny_llm} & Commercialized \\ 
\midrule
\textbf{Phoenix} & Sanctuary AI & Humanoid & Collaborative work, service tasks & Advanced cognitive AI & Human-like behavior and natural interaction & Prototype \\ 
\midrule
\textbf{Optimus Gen 2} & Tesla & Humanoid & Industrial and repetitive tasks & Tesla proprietary AI stack & Functional persona & Prototype / Pre-production \\ 
\midrule
\textbf{Figure 02} & Figure AI & Humanoid & Industrial automation and manufacturing & OpenAI, NVIDIA partnership & Task-oriented, dexterous persona \cite{openai2024figure} & Prototype \\ 
\midrule
\textbf{Astro} & Amazon & Functional / abstract & Home assistant, security monitoring & Alexa, SLAM navigation, smart mobility & Functional extension of Alexa ecosystem & Commercialized \\ 
\midrule
\textbf{Aibo} & Sony & Zoomorphic (dog) & Emotional companionship & Adaptive personality AI, facial recognition & Pet-like, evolving persona & Commercialized \\ 
\midrule
\textbf{Lovot} & Groove X & Zoomorphic (abstract) & Emotional companionship & AI emotion engine, multimodal sensor fusion & Affection-seeking, comforting persona & Commercialized \\ 
\bottomrule
\end{tabular}}
\end{table*}

\subsubsection{Quadrant IV: Vertical Market Companions (Therapeutic Persona)}
\begin{itemize}
    \item \textbf{Elderly Care (Proactive Coach Persona):}
        \begin{itemize}
            \item \textbf{ElliQ:} A proactive desktop companion designed for seniors. It does not wait for commands but actively initiates conversations, suggests activities (e.g., walking, hydration), tracks wellness, and connects users to family or online communities (e.g., Bingo). Its persona is a friendly, supportive ``coach,'' and studies confirm its utility in aiding daily life \cite{xiong2024elliq}.
        \end{itemize}

    \item \textbf{Autism \& Special Needs (Therapeutic Mediator Persona):}
        \begin{itemize}
            \item \textbf{QTrobot (LuxAI):} An expressive social robot designed for ASD therapy. Its predictable, non-judgmental persona reduces anxiety. It functions as a ``social mediator'' in a triangular relationship, where the robot engages the child, who then practices the same skill (e.g., eye contact) with the human therapist, facilitating generalization \cite{costa2024qtrobot}.
            \item \textbf{Milo (RoboKind):} A humanoid robot that teaches social and emotional skills using a specialized curriculum, leveraging its emotional engine and NLP to model human facial expressions.
        \end{itemize}
\end{itemize}

\subsection{In-Depth Analysis of Core Challenges}
The challenges of deploying robust embodied AI are being rigorously assessed, with new frameworks like EmbodiedBench \cite{embodiedbench2025} being developed to benchmark MLLM performance in these complex, interactive scenarios.

\subsubsection{Technical Barriers}
\begin{itemize}
    \item \textbf{Latency:} Real-time, natural conversation is highly sensitive to cloud-based LLM latency. Edge computing and model optimization are critical research areas.
    \item \textbf{Hallucinations \& Context Deviation:} LLMs generating factually incorrect (but plausible) information is extremely dangerous in high-stakes (e.g., medical) applications. RAG is a primary mitigation strategy.
    \item \textbf{Symbol Grounding Problem:} The fundamental challenge of connecting abstract LLM symbols (the word ``apple'') to physical sensor data (a red, round object). This asymmetry (can talk, cannot ``understand'') is a key focus for VLM (Vision-Language Model) and VLA (Vision-Language-Action) model research \cite{ma2024vla_survey}.
\end{itemize}

\subsubsection{Privacy and Security}
\begin{itemize}
    \item \textbf{Multi-Dimensional Threat:} Robots are ``data gathering terminals'' with cameras, mics, and LIDAR, capable of collecting sensitive data (habits, health, finances) from private spaces (bedrooms).
    \item \textbf{User Psychology:} Users exhibit ``privacy resignation'' (feeling collection is inevitable) but also extreme discomfort with ``data inference'' (the robot ``knowing'' things not explicitly told) \cite{chmielewski2024privacy}.
    \item \textbf{Mitigation Strategy:} A ``privacy-by-design'' approach is mandatory, emphasizing on-device (edge) processing, strong encryption, data anonymization, and transparent user controls \cite{chmielewski2024privacy}.
\end{itemize}

\subsubsection{Ethical and Legal Frameworks}
\begin{itemize}
    \item \textbf{Liability and Accountability:} The ambiguity of who is responsible (user, manufacturer, software developer) if an AI provides harmful medical advice or causes physical damage is a primary barrier to commercialization \cite{kaminski2024liability}.
    \item \textbf{Algorithmic Bias:} Biased training data (e.g., underrepresentation in medical data) can lead to discriminatory or unfair outcomes, which is highly dangerous in diagnostics.
    \item \textbf{Emotional Deception and Dependency:} The ethics of fostering emotional bonds, especially with vulnerable populations (children, elderly), and the risk of substituting robotic care for necessary human care.
    \item \textbf{HIPAA Compliance:} A major challenge for US healthcare. Standard consent does not cover data ``reuse'' for AI training, and ``de-identified'' data faces a high risk of ``re-identification,'' requiring updates to HIPAA safety rules for the AI era \cite{miller2024hipaa_ai}.
\end{itemize}

\subsubsection{Economic and Market Barriers}
\begin{itemize}
    \item \textbf{High Cost:} Advanced hardware R\&D and manufacturing costs make products prohibitively expensive for the mass market.
    \item \textbf{Unclear Value Proposition:} For general-purpose robots, the convenience offered often does not yet justify the high price tag.
    \item \textbf{Expectation Management:} A significant gap exists between sci-fi portrayals and current technological reality, leading to user disappointment \cite{kang2025uncanny_llm}.
\end{itemize}

\subsection{Future Trajectory and Stakeholder Recommendations}

\subsubsection{Key Development Trends}
\begin{itemize}
    \item \textbf{Proactive Intelligence:} Shifting from passive command-execution to proactively anticipating user needs based on learned patterns and real-time context \cite{xiong2024elliq}.
    \item \textbf{Hyper-Personalization:} Using long-term memory and RLHF (Reinforcement Learning from Human Feedback) to develop unique interaction styles for each family member.
    \item \textbf{Enhanced Emotional Intelligence:} Finer-grained understanding of human emotion for more sincere, natural interactions.
\end{itemize}

\subsubsection{Functional Fusion and Ecosystem Integration}
\begin{itemize}
    \item \textbf{Fusion:} The long-term direction is a hybrid model, blending physical assistance (industrial-grade dexterity \cite{openai2024figure}) with social/emotional support.
    \item \textbf{Ecosystem (IoT):} Robots will become the central hub for the smart home, coordinating other IoT devices (lights, security) via standards like Matter.
    \item \textbf{Ecosystem (Telehealth):} Robots will act as ``health-bots-in-the-home,'' serving as mediators for virtual doctor visits, monitoring vital signs, and collecting daily health data.
\end{itemize}

\begin{table*}[t]
\centering
\small
\caption{Major Challenges of LLM-driven and Embodied AI Systems with Root Causes and Mitigation Strategies. The table categorizes the key technical, ethical, and economic challenges, explains underlying causes, and summarizes representative mitigation directions.}
\label{tab:robotics_challenges}
\resizebox{\textwidth}{!}{
\begin{tabular}{@{}p{2.5cm}p{3.8cm}p{5cm}p{5cm}@{}}
\toprule
\textbf{Challenge Category} & \textbf{Specific Challenge} & \textbf{Root Cause Analysis} & \textbf{Mitigation Strategies} \\ 
\midrule
\textbf{Technical} 
& Hallucination & LLMs generate text probabilistically without factual verification mechanisms & Retrieval-Augmented Generation (RAG); integration with verified knowledge bases \\ 
& Latency & Computational overhead of large models running on cloud infrastructure & Edge computing; model quantization and optimization; hardware acceleration \\ 
& Symbol Grounding & Disconnection between linguistic symbols and real-world perception & Vision-Language-Action (VLA) models; multimodal training data \cite{ma2024vla_survey} \\ 
\midrule
\textbf{Privacy \& Security} 
& Intrusive Data Collection & Robots depend on continuous perception of users and environments & Privacy-by-design; data minimization; user transparency and control \cite{chmielewski2024privacy} \\ 
& Data Inference & AI can infer undisclosed sensitive information from multi-source data & Strict data-use policies; user control over inference outputs \cite{chmielewski2024privacy} \\ 
& Security Vulnerabilities & Complex software–hardware ecosystems are targets for cyberattacks & End-to-end encryption; regular security audits; secure update mechanisms \\ 
\midrule
\textbf{Ethical \& Legal} 
& Responsibility Ambiguity & Undefined accountability between user, manufacturer, and developer in case of harm & New robotic legislation; clarified liability framework \cite{kaminski2024liability} \\ 
& Algorithmic Bias & Biased or unbalanced training data amplifies social inequalities & Diverse datasets; bias audits and debiasing algorithms \\ 
& Emotional Dependence & Vulnerable users may develop excessive emotional attachment to robots & Ethical design guidelines; role transparency; avoidance of deceptive behavior \\ 
\midrule
\textbf{Economic} 
& High Hardware Cost & Complex sensors, actuators, and computational units increase production cost & Manufacturing innovation; supply chain optimization; subscription or leasing models \\ 
& Unclear Value Proposition & General-purpose robots lack matching utility for their price point & Focus on high-value verticals; ``utility-first'' design strategy \\ 
& Expectation Gap & Mismatch between public expectations and actual system capabilities & Transparent marketing; realistic expectation management \cite{kang2025uncanny_llm} \\ 
\bottomrule
\end{tabular}}
\end{table*}

\subsubsection{Recommendations for Stakeholders}
\begin{itemize}
    \item \textbf{For Investors:}
        \begin{itemize}
            \item \textbf{Short-term:} Focus on vertical markets with clear ROI (elderly care \cite{xiong2024elliq}, special needs \cite{costa2024qtrobot}), which are the best ``beachheads.''
            \item \textbf{Long-term:} View general-purpose humanoids as a high-risk, high-reward bet. Prioritize companies with breakthroughs in core tech (dexterity, VLA models \cite{ma2024vla_survey}) and clear paths from industrial to consumer markets.
        \end{itemize}
    \item \textbf{For Developers:}
        \begin{itemize}
            \item \textbf{Mass Market Strategy:} Adopt a ``utility-first, persona-second'' approach. Build a market base by solving practical pain points (security, convenience) first.
            \item \textbf{Vertical Market Strategy:} Engage in deep co-design with domain experts (doctors, therapists) and end-users.
            \item \textbf{Universal Principle:} ``Privacy-by-design'' must be non-negotiable. Trust is the core competitive advantage \cite{chmielewski2024privacy}.
        \end{itemize}
    \item \textbf{For Policymakers:}
        \begin{itemize}
            \item \textbf{Legislate Proactively:} Urgently develop clear legal frameworks for liability \cite{kaminski2024liability}, data privacy (e.g., updating HIPAA for AI \cite{miller2024hipaa_ai}), and algorithmic accountability.
            \item \textbf{Set Standards:} Drive industry standards for data security, interoperability, and ethical design (e.g., using benchmarks like \cite{embodiedbench2025}).
            \item \textbf{Educate Public:} Manage societal expectations by fostering rational, public discussion about the technology's true capabilities and limitations.
        \end{itemize}
\end{itemize}

\end{document}